\documentclass{article}

\usepackage{amsmath}
\usepackage{amssymb}
\usepackage{amsthm}
\usepackage{graphicx}
\usepackage{hyperref}
\usepackage{fancyhdr}
\usepackage{latexsym}
\usepackage{authblk}

\usepackage[sorting=ynt]{biblatex}
\hypersetup{colorlinks=true,linkcolor=blue,citecolor=blue}

\theoremstyle{plain}
\newtheorem{theorem}{Theorem}[section]

\theoremstyle{remark}
\newtheorem*{Proof}{Proof}

\theoremstyle{definition}
\newtheorem{remark}{remark}[section]

\theoremstyle{plain}
\newtheorem{coro}{Corollary}[section]

\newcounter{fig}

\newcommand{\D}{\textrm{d}}
\newcommand{\X}{X^{*}}
\newcommand{\I}{I^{*}}
\newcommand{\nfig}{\stepcounter{fig}\arabic{fig}}
\newcommand{\cfig}{\arabic{fig}}
\newcounter{tab}
\newcommand{\ntab}{\stepcounter{tab}\arabic{tab}}

\title{	\textbf{{\huge Emergence of Hopf bifurcation in an extended SIR dynamic}}}

\author[1,2]{Arash Roostaie}
\author[2,1]{Hadi Barzegar}
\author[1,3,*]{Fakhteh Ghanbarnejad}

\affil[1]{Department of Physics, Sharif University of Technology, P.O. Box 11165-9161, Tehran, Iran}
\affil[2]{Department of Mathematical Sciences, Sharif University of Technology, P.O. Box 11165-9161, Tehran, Iran}
\affil[3]{Quantitative Life Sciences (QLS), The Abdus Salam International Centre for Theoretical Physics (ICTP),
Strada Costiera, 11, I-34151 Trieste, Italy}
\affil[*]{fakhteh.ghanbarnejad@gmail.com}

\date{}
\begin{document}
	\maketitle
	\newpage
	
	\tableofcontents
	\newpage
	\label{IM2}
	
	\section{Abstract}
	In this paper, the SIR dynamics is extended by considering another compartmental which represents hospitalization of the critical cases. So a system of differential equations with four blocks is considered when there is intensive care unit (ICU) to cure critical cases. Outgoing rate of survived infected individuals is divided into $nI$ and $\frac{bI}{I+b}$. The second term represents the rate of critical cases who enter ICUs. It is proved that there are forward, backward and Hopf bifurcations in different regimes of parameters.  
	
	\section{Introduction}
	In the article by Kermack and Mckendrick \cite{kermack1927contribution}, the first susceptible-infected-recovered(SIR) model was introduced in order to simulate and predict a diseases spreading phenomena and its epidemic. In this standard model, there are three compartments labeled as $S$, $I$ and $R$ which indicate number of (or percentage of) susceptible, infectious and recovered individuals and the transitions between these blocks happen with the constant parameters $\beta$, the infection rate, and $\gamma$, the recovery rate.
	
	After publishing of this article, the parameters $\beta$ and $\gamma$ have been changed in various works and also some new blocks have been introduced and simulated in order to achieve more accurate models for better prediction of the behavior of different epidemics. For instance, $\beta$ has been considered as $kI^{p-1}S^{q-1}$,$\beta^{aS+bI+cR}$ and $\mu e^{-mI}$, in the articles \cite{[20]},\cite{[16]},\cite{[17]}, respectively. In these articles, it has been tried to model behavior of people and strategies of government with considering $\beta$ as a function of other blocks. 
	
	In order to consider capacity of the heath-care system in a country, $\gamma$ has been changed in articles too. For instance, $\gamma I$ was been considered as:
	\begin{align}
		\gamma I=
		\begin{cases}
		0\qquad &I=0\\
		rI\qquad &I>0
		\end{cases}
	\end{align}
	\begin{align}
		\gamma I=
		\begin{cases}
		rI\qquad &0\le I\le I_0\\
		k\qquad &I_0<I
		\end{cases}
	\end{align}	
	\begin{align}
		\gamma I=(\mu_0+(\mu_1-\mu_0)\frac{b}{b+I})I
	\end{align}	
	 in the articles \cite{[22]},\cite{[23]} and \cite{[24]} repectively.
	 
	In our study, an improved model, which considers the effect of capacity ICU (Intensive Care Unit) in a hospital, will be discussed. To model this effect, a new variable $H$ will be defined which expresses the number of individuals in ICU. So there is a new block in differential equations of SIR model. Fig. \ref
	{Fig:DiagramModel} represents a schematic of this model and the dynamics' equations are expressed as follow:
	\begin{align}
		\begin{cases}
		\frac{\D S}{\D t}=A -dS-\beta I S\\\\
		\frac {\D I}{\D t}=\beta I S-dI-\alpha I-nI-\mu(I)\\\\
		\frac {\D H}{\D t}=\mu(I)-dH-n'H\\\\
		\frac {\D R}{\D t}=nI+n'H-dR
		\end{cases}	
		\label{ASIHR}	
	\end{align} 
	\begin{center} \label{Fig:DiagramModel}
		\includegraphics[width=.7\linewidth]{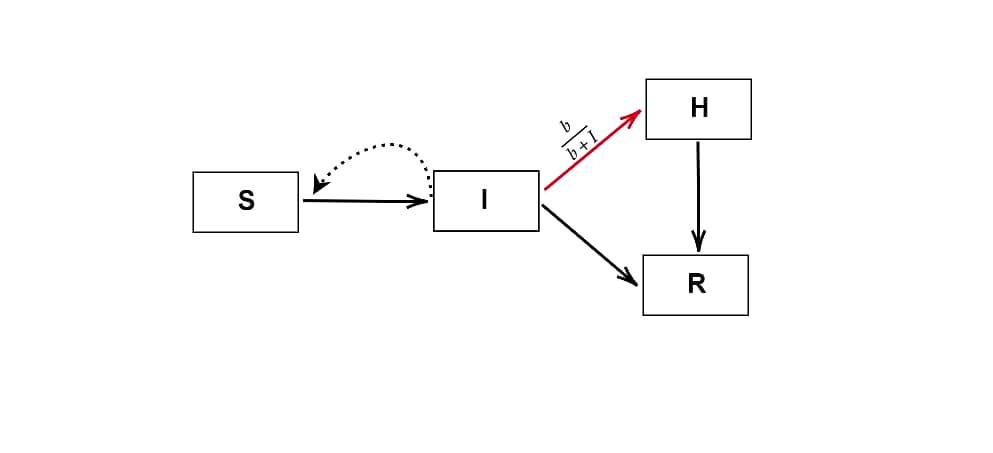}\\
		\textbf{Figure \nfig:} A schematic of our model. Please note that the natural birth and death are not shown here and all the quantities and parameters of the model are listed in Table 1 and 2. 
	\end{center}
	In the equations \eqref{ASIHR}, A is birth rate, $d$ natural death rate and $\alpha$ is death rate caused by disease and $\beta$ is the incidence rate. In this new model, the individuals of $I$, that will be survived, will be divided in two groups. The individuals of first group will be cured directly and they will not need to use special medical treatments of health system such as ICU so they will be transported in block $R$ with a rate $n$. In other hand, the individuals of second group need to get special medical treatments and they must be cured in hospitals. These individuals will be transported in a new block $H$  with a rate $\mu(I)$. The individuals of $H$ will be recovered with the rate $n'$.
	\begin{remark}
		Obviously, the parameters $n$ and $n'$ must be functions of $I$ and $H$ and there must be a connection between $I$ and $H$, for example $\alpha$ can be a function of $H$, but we will ignore these effects to simplify of equations. So $n$,$n'$ and $\alpha$ will be considered as constants.  
	\end{remark}
	A function, that can model $\mu(I)$ in real situations, can be expressed as follow, where $b$ indicates the number of beds in ICU:
	\begin{align*}
		\mu_1(I)=
		\begin{cases}
		I\qquad I\le b\\\\
		b\qquad b\le I
		\end{cases}
	\end{align*}
	At first, when the number of infected individuals $I$ is less than the number of beds $b$,  the rate of transport to ICU will be $I$, for example, there are 100 beds and 50 infected people, so exactly 50 individuals will be transported in ICU. On the other hand, when the number of  infected individuals $I$ is more than the number of beds $b$, the rate of transport to ICU will be independent on the number of infected individuals and will be $b$, for example, there are 150 infected people and 100 beds, so 100 infected people will be transported in ICU.
	
	But there is a big problem about this function. The function is obviously undifferentiable at the point $I=b$. So to simplify the solving of differential equations, we will choose another function, which is similar to this function and will be smooth. A good choice can be expressed as follows:
	\begin{align*}
		\mu_2(I)=\frac {bI}{I+b}
	\end{align*}
	Obviously, in the limits $I \rightarrow0$ and $I \rightarrow \infty$, we have $\mu_2(I)\rightarrow\mu_1(I)$ and $\frac {\D\mu_2}{\D I}\rightarrow\frac {\D\mu_1}{\D I}$ will be defined as follow:
	\begin{align*}
		\mu(I):=\frac {bI}{I+b}
	\end{align*}
	And the differential equations will be expressed as follow:
	\begin{align}
		\begin{cases}
		\frac{\D S}{\D t}=A -dS-\beta I S\\\\
		\frac {\D I}{\D t}=\beta I S-dI-\alpha I-nI-\frac {bI}{I+b}\\\\
		\frac {\D H}{\D t}=\frac {bI}{I+b}-dH-n'H\\\\
		\frac {\D R}{\D t}=nI+n'H-dR
		\end{cases}	
		\label{aSIHR}	
	\end{align}
	\\
	\begin{center}
		\begin{tabular}{|c|c|}
			\hline
			\textbf{quantities}&\textbf{definition}\\
			\hline
			$S(t)$&susceptible individuals\\
			\hline
			$I(t)$&infected individuals\\
			\hline
			$H(t)$&individuals of ICU\\
			\hline
			$R(t)$&recovered individuals\\
			\hline
		\end{tabular}
		\begin{tabular}{|c|c|}
			\hline
			\textbf{parameter}&\textbf{definition}\\
			\hline
			$\beta$&infectious rate\\
			\hline
			$b$&number of beds\\
			\hline
			$A$&birth rate\\
			\hline
			$d$&natural death rate\\
			\hline
			$\alpha$&death rate caused by infectious\\
			\hline
			$n$&natural recovery rate\\
			\hline
			$n'$&recovery rate  of individuals of ICU\\
			\hline
		\end{tabular}
	\end{center}
		\textbf{Table \ntab,\ntab:} List of all quantities and parameters of the model. Dimension of all parameters except $b$ is $T^{-1}$. $b$ and quantities are dimensionless.\\
		
	In the system \eqref{aSIHR}, the first two equations are independent of the other equations, so to analyse the system \eqref{aSIHR}, such as finding the fixed points and bifurcations, just these equations will be considered. Thus the essential equations are:
	 \begin{align}
		 \begin{cases}
		 \frac{\D S}{\D t}=A -dS-\beta I S\\\\
		 \frac {\D I}{\D t}=\beta I S-\delta I-\frac {bI}{I+b}
		 \end{cases}	
		 \label{SIHR}	
	 \end{align}
	where $\delta$ is defined as follow:
	\begin{align*}
	 	\delta:=d+\alpha+n
	\end{align*}
	\section{Fixed Points}
	If one presents the system \eqref{SIHR} as $\dot{X}=f(X)$, to find fixed points, $f(X^{*})=0$ must be solved, so:
	 \begin{align}
		\begin{cases}
		A -dS-\beta I S=0\\\\
		\beta I S-\delta I-\frac {bI}{I+b}=0
		\end{cases}	
	\end{align}	
	With solving second equation, one can find:
	 \begin{align*}
	 	\begin{cases}
	 	I=0\\
	 	or\\
	 	S=\frac {1}{\beta} (\delta+\frac {b}{I+b})
	 	\end{cases}
	 \end{align*}
	Now with combining the first case and the first equation, one fixed point can be found as follow:
	\begin{align*}
		E_0=(\frac {A}{d},0)
	\end{align*}
	Which is the \emph{disease free equilibrium(DFE)}.

	With combining the second case and the first equation, the following quadratic equation can be found:
	\begin{align}
		\label{quad0}
		\delta I^2+(b(\delta+1)+\frac {\delta d}{\beta}-A)I+b(\frac {d}{\beta}(\delta+1)-A)=0
	\end{align}
	As we see in the next sections, it will be useful to define $R_0$,\emph{basic reproduction number}, as follow:
	\begin{align*}
		R_0:=\frac {\beta A}{d(\delta+1)}
	\end{align*}
	Now with this definition, the equation \eqref{quad0} can be rewritten as follow:
	\begin{align}
		P(I)=c_2I^2+c_1I+c_0=0	
		\label{quad}
	\end{align}
	With following coefficients:
	\begin{align*}
		\begin{cases}
		c_2=\delta\\\\
		c_1=b(\delta+1)+\frac {\delta d}{\beta}-A\\\\
		c_0=\frac {bd(\delta+1)}{\beta}(1-R_0)
		\end{cases}
	\end{align*}

	So there will be following theorems about other fixed points $X^{*}\ne E_0$ of system \eqref{SIHR}.
	\begin{theorem}
		There is exactly one fixed point $\X$ if $R_0>1$.
	\end{theorem}
	\begin{Proof}
		It is obvious:
		\begin{align*}
			R_0>1\Rightarrow\frac {bd(\delta+1)}{\beta}(1-R_0)<0
		\end{align*}
		so the quadratic equation \eqref{quad} have two solutions $I_1$ and $I_2$ with the condition $I_1I_2<0$. But in the system \eqref{SIHR}, $I$ must be non-negative, so there will be just one permissible solution for the quadratic equation \eqref{quad} and this proves this theorem.$\Box$
	\end{Proof}
	\begin{theorem}
		If $R_0<\frac {\delta}{(\delta+1)}$, there is no fixed point $\X$.
	\end{theorem}
	\begin{Proof}
	it is obvious that $R_0<1$ so $c_0>0$ and we can drive from the assumption that:
	\begin{align*}
		R_0<\frac {\delta}{(\delta+1)} \Rightarrow &\frac {A\beta}{d(\delta+1)}<\frac {\delta}{(\delta+1)} \\
		\Rightarrow &\frac {A\beta}{d}<\delta\\
		\Rightarrow &0< \frac {d\delta}{\beta}-A \quad \& \quad 0<b(\delta+1)\\
		 \Rightarrow &0<b(\delta+1)+\frac {\delta d}{\beta}-A\\
		 \Rightarrow &0<c_1
	\end{align*}
	So firstly, if there are real solutions for quadratic equation \eqref{quad}, both of them must be negative ($c_1>0 \:\&\:  c_0>0$). On the other hand there are no real solutions for this equation. So the theorem has been proved.$\Box$	
	\end{Proof}
	\begin{theorem}
			If $\frac {\delta}{(\delta+1)}<R_0<1$, there will be following cases:
			\begin{description}
				\item[\boldmath{$b>\frac{d}{\beta}(\frac{1}{\delta+1})$}:] there is no fixed point $\X$.
				\item[\boldmath{$b<\frac{d}{\beta}(\frac{1}{\delta+1})$}:] there are two fixed points $\X$ for $R_0$ close enough to 1 .
			\end{description}
	\end{theorem}
	\begin{Proof}
		If $R_0<1$, it is obvious for two cases that $0<c_0$.  Now we can drive following result for first case:
		\begin{align*}
			&\frac{d}{\beta}(\frac{1}{\delta+1})<b\Rightarrow \frac{d}{\beta}<b(1+\delta)\\
			\Rightarrow&\frac{(1+\delta)d}{\beta}<b(1+\delta)+\delta\frac{d}{\beta}\\
			\Rightarrow&\frac{(1+\delta)d}{\beta}-A<b(1+\delta)+\delta\frac{d}{\beta}-A\\
			\Rightarrow&\frac{(1+\delta)d}{\beta}(1-\frac{\beta A}{d(\delta+1)})<c_1\\
			\Rightarrow&\frac{(1+\delta)d}{\beta}(1-R_0)<c_1\: \& \: 0<1-R_0\\
			\Rightarrow &0<c_1
		\end{align*}
		Because $c_1>0$ and $c_0>0$, the proof of this case will be similar to the previous theorem. So there is no fixed point $\X$ in this case.
		
		The second case is a bit more complicated. At first, the new parameter $\epsilon$ will be defined as follow:
		\begin{align*}
			\epsilon:=\frac{d(\delta+1)}{\beta}-A
		\end{align*}
		$\epsilon$ will be positive, because:
		\begin{align*}
			R_0<1\Rightarrow\frac{\beta A}{d(\delta+1)}&<1\Rightarrow A<\frac{d(\delta+1)}{\beta}\\
					0&<\epsilon\\
		\end{align*}
		and the following properties are obvious:
		\begin{align*}
			&A=\frac{d(\delta+1)}{\beta}-\epsilon\\
			&\epsilon\rightarrow0^{+} \Leftrightarrow R_0 \rightarrow1^{-}
		\end{align*}
		Now we claim, if $\epsilon$ is small enough, $c_1<0$:
		\begin{align*}
			c_1<0 \Leftrightarrow &b(\delta+1)+\frac {\delta d}{\beta}-A<0\\
			\Leftrightarrow &0<A-b(\delta+1)-\frac {\delta d}{\beta}\\
			\Leftrightarrow &0<\frac{d(\delta+1)}{\beta}-\epsilon-b(\delta+1)-\frac {\delta d}{\beta}\\
			\Leftrightarrow &\epsilon<\frac{d}{\beta}-b(\delta+1)\\
			\Leftrightarrow &\epsilon<(\delta+1)(\frac{d}{\beta(\delta+1)}-b)
		\end{align*} 
		But because in the second case $b<\frac{d}{\beta(\delta+1)}$, the right side of the last inequality is positive, so $\epsilon$ can be found. So in this case:
		\begin{align*}
			c_1<0 \Leftrightarrow \epsilon<(\delta+1)(\frac{d}{\beta(\delta+1)}-b)
		\end{align*}
		Now with considering the properties of $\epsilon$, which is mentioned above, we can find, that the right inequality is equal to the condition, $R_0$ close enough to $1$.
		
		Now it is obvious, when $c_1<0$ and $c_2>0$, $I_{min}:= arg min\: P(I)$ is positive and when $R_0=1$, there are two solutions for quadratic equation \eqref{quad} $I_1=0$ and $I_2>0$. So when $R_0=1$, $P(I_{min})<0$. Now with decreasing $R_0$, when $R_0$ is close enough to $1$ and the conditions $c_1<0$ and $P(I_{min})<0$ are hold, there must be two positive solutions for the quadratic equation \eqref{quad}.(the condition $P(I_{min})<0$ can be hold because of continuously). So the second case of theorem can be proved.$\Box$
	\end{Proof}
	\begin{coro}
		When $R_0<1$, $E_0$ will be a unique fixed point if $b>\frac{d}{(\delta+1)\beta}$ and there will be two other fixed points $\X$  if $b<\frac{d}{(\delta+1)\beta}$.
	\end{coro}
	\section{Stability of Fixed Points}
	Firstly, to analyze the stability of fixed points, $Df$ must be determined. $Df$ will be expressed as follow:
	\begin{align}
	Df(X)=
		\begin{pmatrix}
		-d-\beta I&-\beta S\\\\
		\beta I& \beta S-\delta-\frac{b^2}{(I+b)^2}
		\end{pmatrix}
	\end{align}
	Now $Df$ at the point $E_0$ can be expressed as follow:
	\begin{align}
		Df(E_0)=
		\begin{pmatrix}
		-d&-\frac {\beta A}{d}\\\\
		0&\frac{\beta A}{d}-(\delta+1) 
		\end{pmatrix}
		=
		\begin{pmatrix}
		-d&-\frac {\beta A}{d}\\\\
		0&(\delta+1)(R_0-1) 
		\end{pmatrix}
	\end{align}
	
	Now there will be following theorem:
	\begin{theorem}
		$E_0$ will be a stable node if $R_0<1$ or will be a saddle point if $R_0>1$.
	\end{theorem}
	\begin{Proof}
		Because $Df(E_0)$ is a upper triangular matrix, its eigenvalues will be $-d$ and $(\delta+1)(R_0-1)$. Now if $R_0<1$, there are two negative eigenvalues and $E_0$ will be a stable node and if $R_0>1$, there is a negative and a positive eigenvalue and $E_0$ will be a saddle point.$\Box$
	\end{Proof}
	
	$Df$ at other fixed points $\X$, when they exist, can be expressed as follow:
	\begin{align}
		Df(\X)=
		\begin{pmatrix}
		-d-\beta I&-(\delta+\frac{b}{I+b})\\\\
		\beta I&\frac{b}{I+b} -\frac{b^2}{(I+b)^2}
		\end{pmatrix}
		=
		\begin{pmatrix}
		-d-\beta I&-(\delta+\frac{b}{I+b})\\\\
		\beta I&\frac{bI}{(I+b)^2}
		\end{pmatrix}
	\end{align}
	Now, the trace of $Df(\X)$, which is the sum of eigenvalues of $Df(\X)$, will be expressed as follow:
	\begin{align*}
		\lambda_1+\lambda_2(I)= tr(Df(\X))=\frac{Ib}{(I+b)^2}-d-\beta I\\
	\end{align*}
	For determinant of $Df(\X)$, which is the product of eigenvalues of $Df(\X)$, one can drive the following equations:
	\begin{align*}
		\lambda_1\lambda_2(I)=det(Df(\X))&=\beta I (\delta+\frac{b}{I+b})-\frac{dbI}{(I+b)^2}-\frac{b\beta I^2}{(I+b)^2}\\
		&=\frac{I}{(I+b)^2}(\beta(\delta(I+b)^2+b(I+b))-db-\beta b I)\\
		&=\frac{\beta I}{(I+b)^2}(\delta I^2+2b\delta I+b^2((\delta+1)-\frac{d}{b\beta}))
	\end{align*}
	Now, there will be following theorems about stability of fixed points $\X$:
	\begin{theorem}
		When $R_0>1$, the unique fixed point $\X$ will be stable if $\frac{1}{4}<d$ and $\frac {d}
		{(\delta+1)\beta}< b$.	
	\end{theorem}
	\begin{Proof}
		Firstly, the necessary condition, which must be satisfied in both of real and complex eigenvalues cases, is that, $\lambda_1+\lambda_2<0$, i.e:
		\begin{align*}
			\frac{Ib}{(I+b)^2}-d-\beta I<0
		\end{align*} 
		Now, it is easy to see, that the function $\frac{Ib}{(I+b)^2}$ is less than or equal to $\frac{1}{4}$ for nonnegative $I$. And as we saw in the theorem, there will be a fixed point with $I \in (0,+\infty)$ when $R_0>1$. So the above inequality must be satisfied for $\forall I \in (0,+\infty)$. So:
		\begin{align*}
			\frac {1}{4}<d &\Rightarrow -d<-\frac {1}{4}\\
			\Rightarrow\frac {Ib}{(I+b)^2}-d&<\frac {Ib}{(I+b)^2}-\frac {1}{4}<0\\
			\Rightarrow\frac {Ib}{(I+b)^2}&-d-\beta I<0
		\end{align*} 
		But when the eigenvalues of $Df(\X)$ are real, an other condition, which must be satisfied, is that, $\lambda_1\lambda_2>0$, i.e:
		\begin{align*}
			\frac{\beta I}{(I+b)^2}(\delta I^2+2b\delta I+b^2((\delta+1)-\frac{d}{b\beta}))>0 \qquad \forall I \in (0,+\infty)
		\end{align*}
		It is easy to see, that the above inequality will be satisfied in this interval, if and only if:
		\begin{align*}
			b^2((\delta+1)-\frac{d}{b\beta})&>0\Rightarrow\\
			(\delta+1)-\frac{d}{b\beta}&>0\Rightarrow\\
			\frac{d}{(\delta+1)\beta}&< b
		\end{align*}
		And this completes the proof of this theorem.$\Box$
	\end{Proof}
	\section{Bifurcations}
	In this subsection will be shown that the system \eqref{SIHR} undergoes three kinds of bifurcation, forward and backward bifurcation and Hopf bifurcation.
	\subsection{forward \& backward bifurcation}
	\begin{theorem}[\textbf{forward bifurcation}]
		When $R_0=1$, the system \eqref{SIHR} undergoes a forward bifurcation if $\frac{1}{4}<d$ and $\frac{d}{(\delta+1) \beta}<b$.
	\end{theorem}
	\begin{Proof}
		According to the theorems of the section 3and 4, when $\frac{d}{(\delta+1) \beta}<b$, $E_0$ is a unique stable fixed point for $R_0<1$ and when $\frac{1}{4}<d$ and $\frac{d}{(\delta+1) \beta}<b$, there will be a unique stable fixed point $\X$ for $R_0>1$ and $E_0$ will be an unstable fixed point.
		So when $\frac{d}{(\delta+1) \beta}<b$ and $\frac{1}{4}<d$ , both of conditions will be satisfied and there will be a forward bifurcation in the system \eqref{SIHR}.$\Box$

		As an example, this theorem can be seen in the following plots:
		\begin{center}
			\includegraphics[width=.49\linewidth]{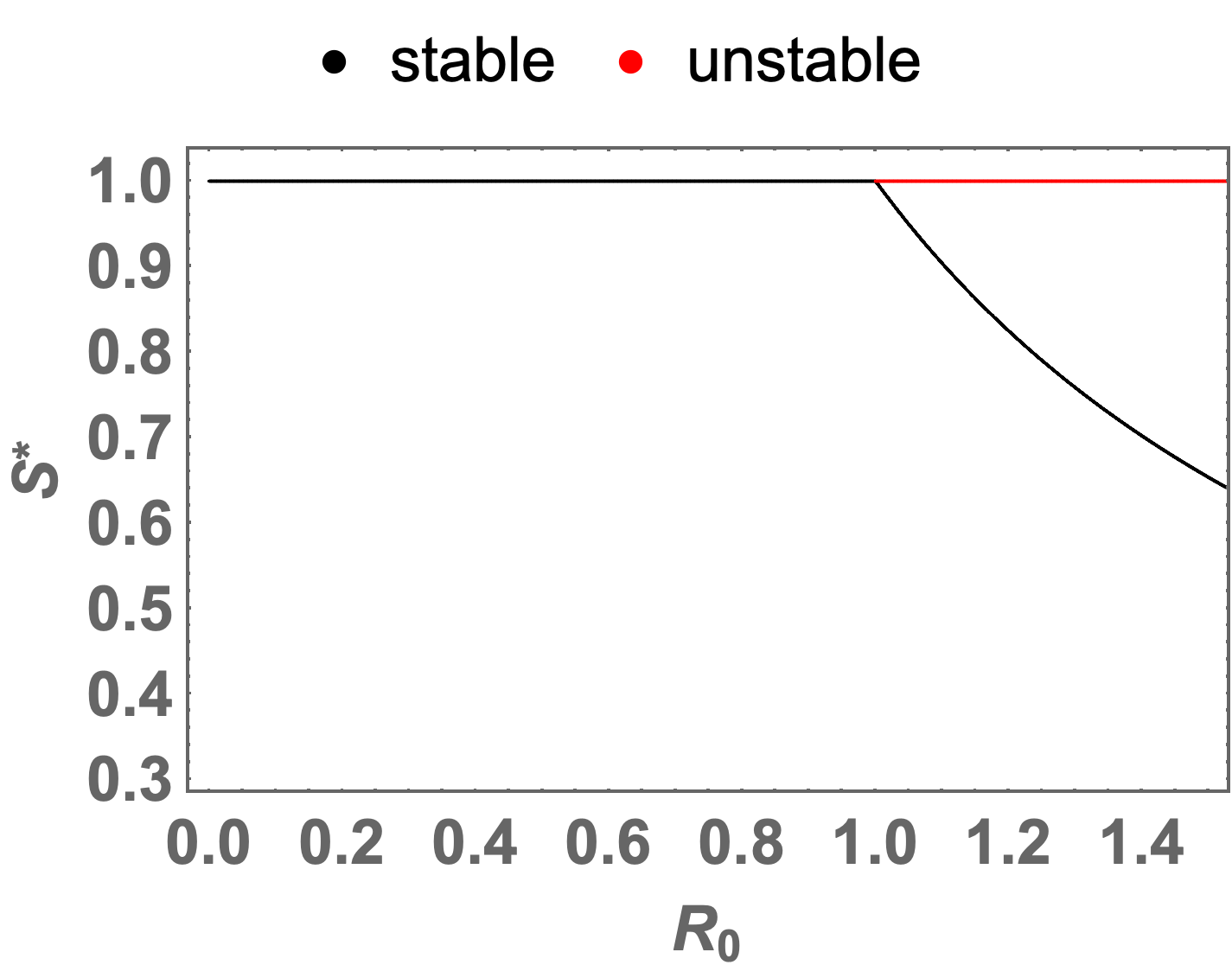}
			\includegraphics[width=.49\linewidth]{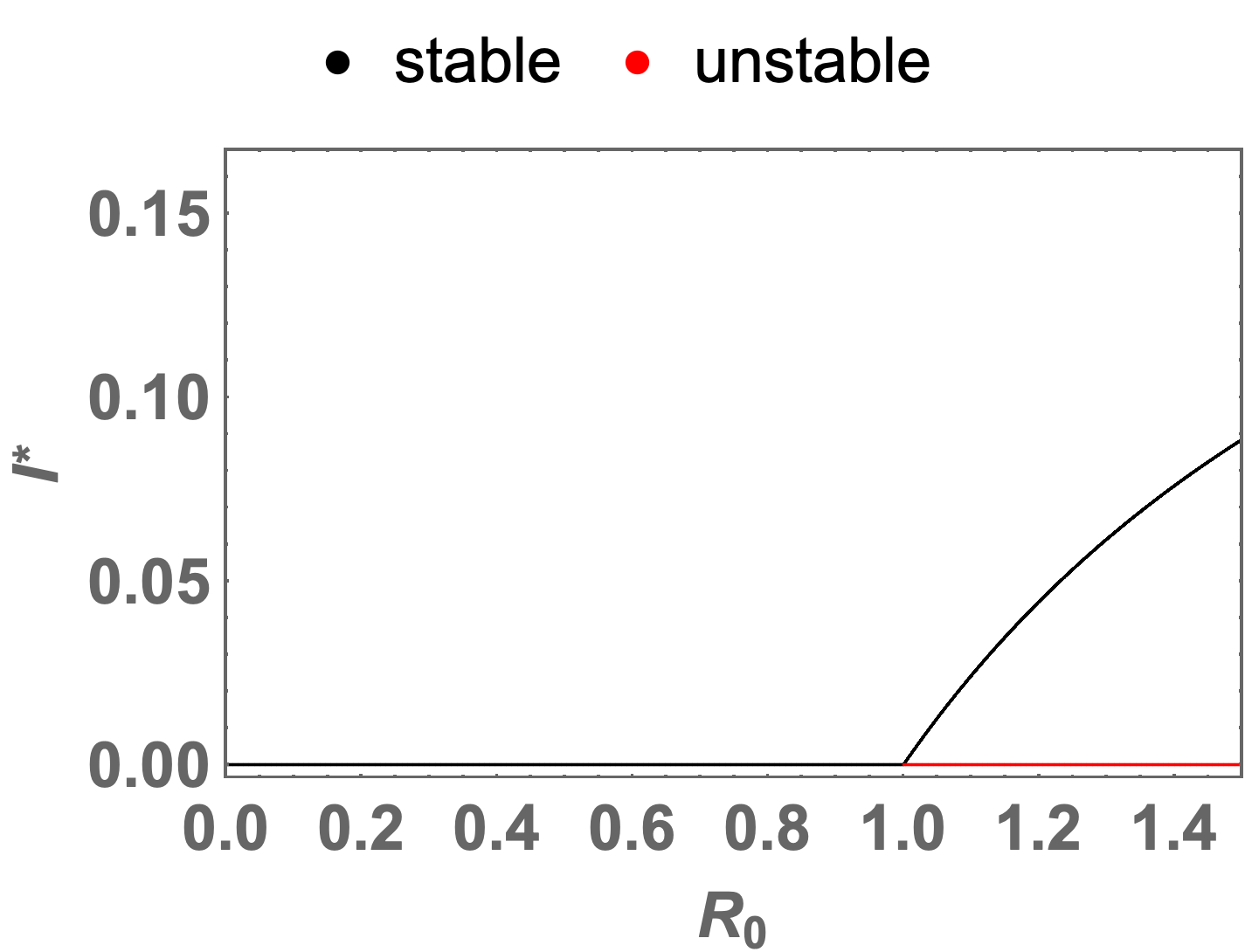}\\
		\end{center}
			\textbf{Figure \nfig:} Simulation of theorem 4.1: plots of $S^*$ and $I^*$, coordinates of fixed points, as a function of $R_0$ where black and red curves indicate stable and unstable fixed points respectively and when the parameters are $b=1,A=1,d=1,\alpha=1,n=1,n'=1$ and $\beta$ changes. The quantities and parameters have been defined in the table 1 and 2 and $R_0$ has been defined in the section 2. 
	\end{Proof}
	\begin{theorem}[\textbf{backward bifurcation}]
	When $R_0=1$, the system \eqref{SIHR} undergoes a backward bifurcation if $\frac{1}{4}<d$ and $b<\frac{d}{(\delta+1) \beta}$.
	\end{theorem}
	\begin{Proof}
		According to the theorems in the section 3 and 4, when $R_0<1$, $E_0$ is always a stable fixed point and there will be two other fixed points $\X$ if $b<\frac{d}{(\delta+1) \beta}$ and $R_0$ close enough to 1. Now we claim that one of these fixed points $\X$ will be a stable node and the other one will be a saddle point. We prove this with \emph{Index Theory}. Firstly, we choose one closed curve $C$ as fallow:
		\begin{center}
			\includegraphics[width=.5\linewidth]{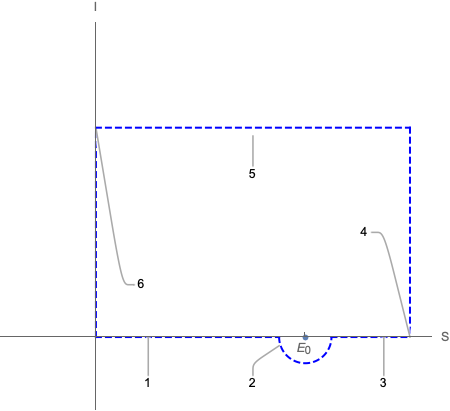}\\
			\textbf{Figure \nfig:} closed curve $C$
		\end{center}
		This curve $C$ has been chosen large enough so when the fixed points $\X$ exist, this curve $C$ will enclose them.(The circle in the fourth quadrant is too small so it encloses just $E_0$ for $R_0<1$).
		
		Now, when $R_0<1$ and small, there is just $E_0$ in the curve $C$, so the index of closed curve $C$ will be as follow:
		\begin{align*}
			I_C=I_0
		\end{align*}
		where $I_0$ is the index of  $E_0$. $E_0$ is a stable node so $I_0=1$ and:
		\begin{align*}
			I_C=1
		\end{align*}
		Now assume that $R_0$ is close enough to 1 and two other fixed points $\X$ exist in the $C$. So $I_C$ will be expressed as follow:
		\begin{align*}
			I_C=I_0+I_1+I_2=1+I_1+I_2
		\end{align*}
		where $I_1$ and $I_2$ are the indexes of other fixed points $\X$.
		So $I_C$ can be considered as a function of $R_0$.
		
		Now we claim that the $I_C(R_0)$ is a continuous function. $I_C(R_0)$ can be rewritten as an integral in the complex plane as follow:
		\begin{align*}
			I_C(R_0)=\frac{1}{2\pi i}\oint_{\tilde{C}} \:\frac{1}{\xi} \:\D\xi 
		\end{align*}
		where $\tilde{C}$ is $f(C)$($C$ has been considered as a closed curve in $\mathbb{C}$ and f as a function $\mathbb{C}\rightarrow\mathbb{C})$. So we can obtain:
		\begin{align*}
			I_C(R_0)=\frac{1}{2\pi i} \sum_{i=1}^{6} \int_0^1 \:\frac{1}{f(g_i(t))}f'(g_i(t))g'_i(t) \:\D t
		\end{align*}
		where $g_i(t)$ are the parameterization of lines and circle of closed curve $C$ in complex plane as it is shown in figure \cfig.
		\\
		Now $f$ and $f'$ are continuous function of $R_0$, so integrands of above integrals are continuous function of $R_0$ and $I_C(R_0)$ will be a continuous function of $R_0$.
		
		Now we know $I_{1,2}=\pm 1$ and $I_C(R_0)$ is continuous, so:
		\begin{align*}
			&\begin{cases}
			I_C(R_0)=1 \qquad &R_0 \in (0,R_0^{(1)})\\\\
			I_C(R_0)=1+I_1+I_2 \qquad &R_0\in (R_0^{(1)},1)\\
			\end{cases}\\\\
			\Rightarrow &I_1+I_2=0\\\\
		  	\Rightarrow &
			\begin{cases}
			I_1=1 \:\&\: I_2=-1\\
			or\\
			I_2=1 \:\&\: I_1=-1\\
			\end{cases}
		\end{align*}
		So one of the fixed points $\X$ must be a saddle point($I=-1$) and the other one must be a stable or an unstable node($I=1$) but $d>\frac{1}{4}$ and $\lambda_1+\lambda_2<0$, so it must be a stable node.
		
		Now when $R_0>1$, $E_0$ is unstable fixed point(saddle point) and there is always an other fixed point $\X_1$ in the first quadrant and a fixed point $\X_2$ near $E_0$ in the circle($\X_2$ is not a permissible fixed point because it exists in fourth quadrant). So in this case $I_0=-1$ and $I_C(R_0)$ is continuous:
		\begin{align*}
			&\begin{cases}
			I_C(R_0)=1 \qquad &R_0 \in (0,1)\\\\
			I_C(R_0)=-1+I_1+I_2 \qquad &R_0\in (1,+\infty)\\
			\end{cases}\\\\
			\Rightarrow &I_1+I_2=2\\\\
			\Rightarrow &
			I_{1,2}=1
		\end{align*}
		So $\X_{1,2}$ are nodes. But similar to the case $R_0<1$, permissible node must be stable, so $\X_1$ will be a stable fixed point.
		
		So there is just a stable fixed point ($E_0$) for $R_0<R_0^{(1)}$ and two stable fixed point($E_0,\X_1$) and an unstable fixed point ($\X_2$) for $R_0^{(1)}<R_0<1$ and a stable fixed point $\X_1$ and an unstable fixed point $E_0$ for $R_0>1$. So there must be a backward bifurcation in $R_0=1$.$\Box$
	\end{Proof}
	As an example, this theorem can be seen in the following plots:
	\begin{center}
		\includegraphics[width=.49\linewidth]{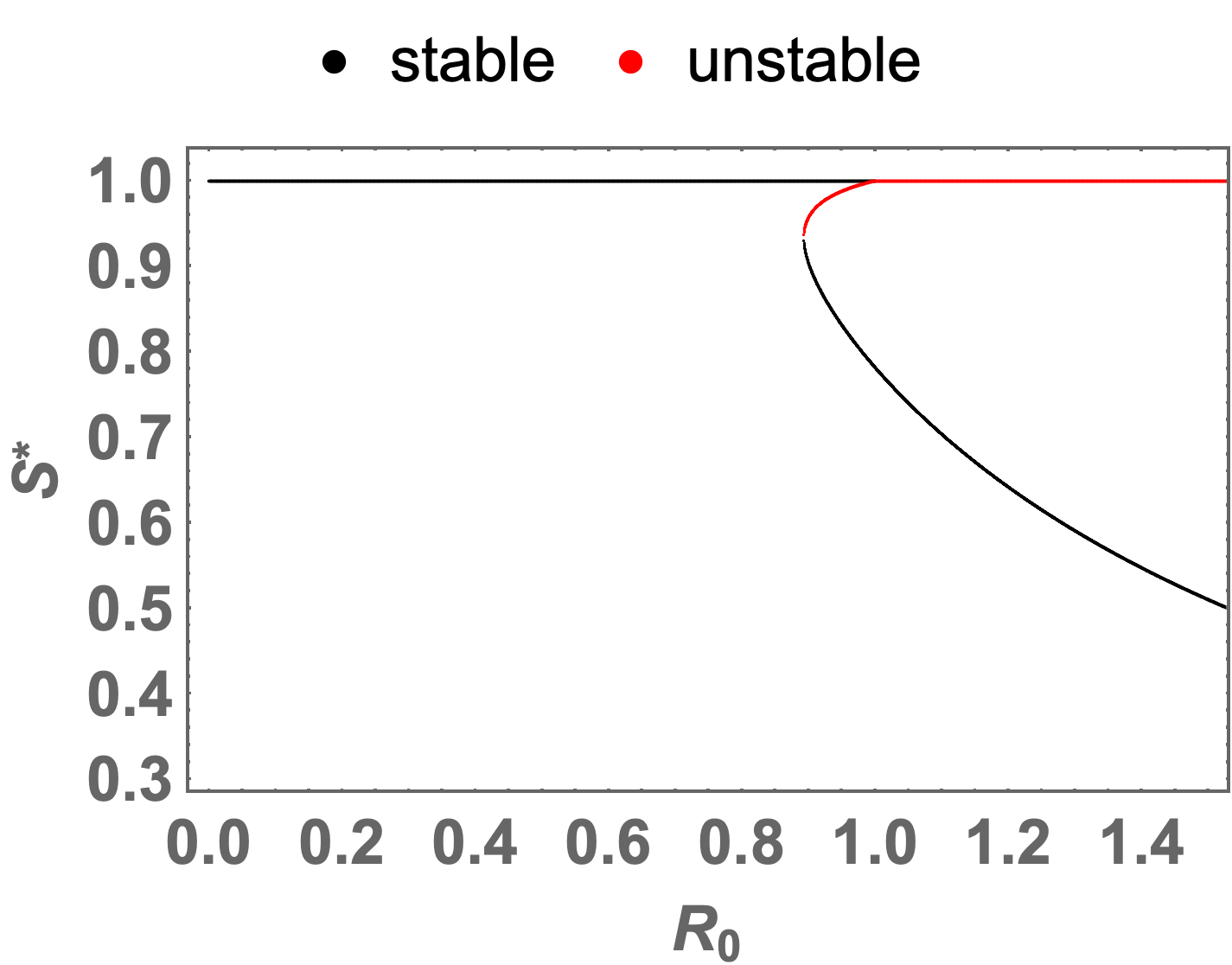}
		\includegraphics[width=.49\linewidth]{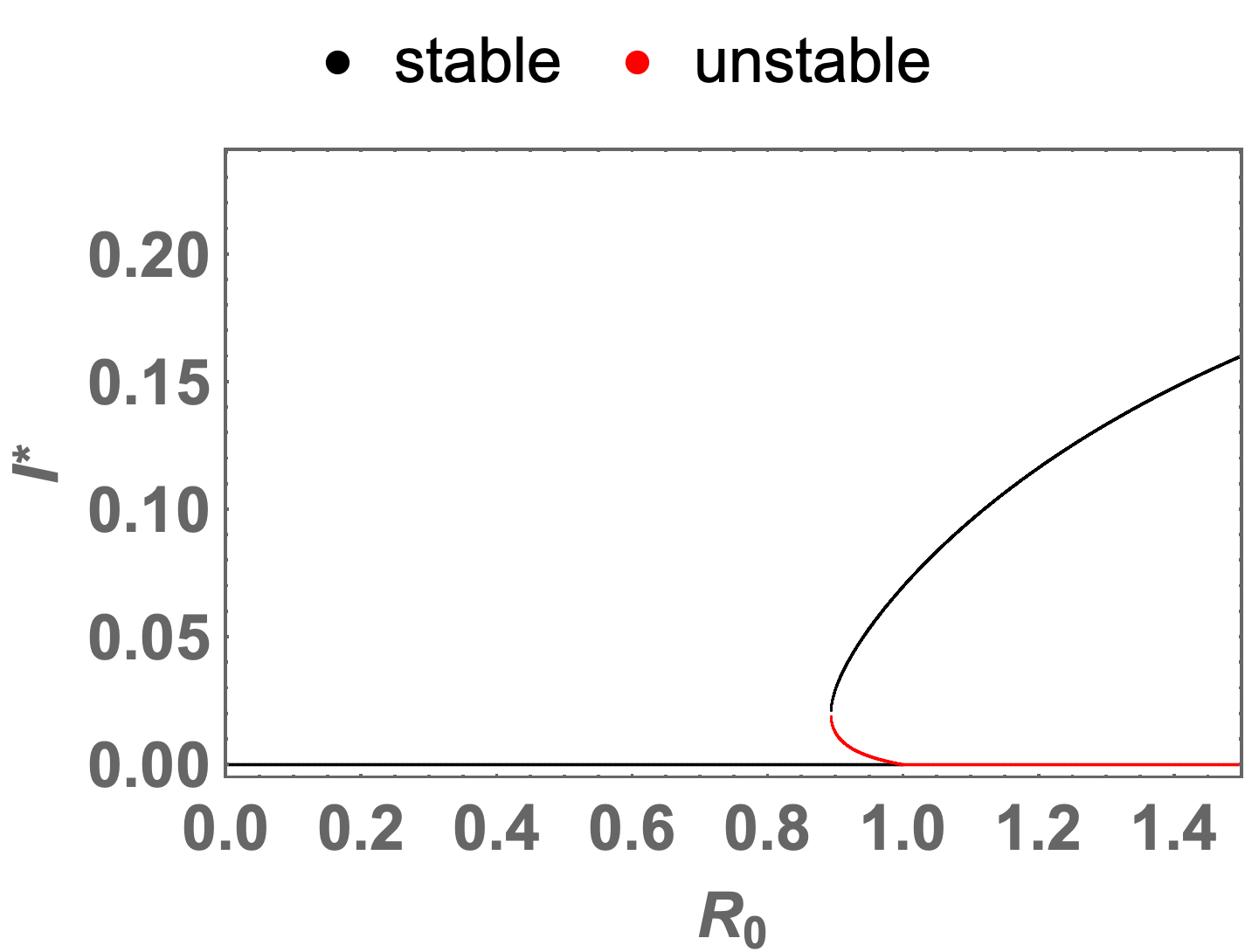}\\
	\end{center}
		\textbf{Figure \nfig:} Simulation of theorem 4.2: plots of $S^*$ and $I^*$,coordinates of fixed points, as a function of $R_0$ where the black and red curves indicate the stable and unstable fixed point respectively and when the parameters are $b=.01,A=1,d=1,\alpha=1,n=1,n'=1$ and $\beta$ changes. The quantities and parameters have been defined in the table 1 and 2 and $R_0$ has been defined in the section 2.
	
	\begin{remark}
		The saddle node bifurcation in backward bifurcation occurs when the quadratic equation \eqref{quad} has just one root , a.e:
		\begin{align}
			\Delta=0\Leftrightarrow c_1^2-4c_2c_0=0
		\end{align}
		Now when $\beta$ is changing and other parameters are constant, $\beta$ can be obtained form the above equation, for example in the simulation of previous theorem, $\beta=0.892857$.
	\end{remark}
	\subsection{Hopf bifurcation}
	\begin{remark}
		In the following theorem, $R_0$ will be as a function of $\beta$ and other parameters will be constant, in other words because $R_0\propto\beta$, there will be no difference between changing of $R_0$ and $\beta$. It will be useful to define $\beta_0$ as follow:
		\begin{align*}
			&\beta_0:=\frac {(\delta+1)d}{A}\\
			\Rightarrow &R_0=1\Leftrightarrow \beta=\beta_0
		\end{align*}
	\end{remark}
	\begin{theorem}[\textbf{Hopf bifurcation}]
		Suppose $\frac{b(2\delta+1)}{2}<A$ and $\frac{d}{(\delta+1)\beta}<b$. Define $\Delta$ and $\beta_{max}$ as follow:
		\begin{align*}
			\Delta:= A- \frac {b(2\delta+1)}{2} \qquad 
			\beta_{max}:=\frac {(\delta+1)d}{\Delta}
		\end{align*}
		With above assumptions, the system \eqref{SIHR} undergoes a Hopf bifurcation for some $\beta \in (\beta_0,\beta_{max})$ or equally for some $R_0 \in (1, \frac {A}{\Delta})$ if $d(1+\frac {(\delta+1)b}{\Delta})<\frac{1}{4}$.
	\end{theorem}
	\begin{Proof}
		Firstly, we will consider the $\lambda_1+\lambda_2(I)$. We have shown:
		\begin{align*}
			\lambda_1+\lambda_2(I)=\frac{Ib}{(I+b)^2}-d-\beta I
		\end{align*}
		First we fix $\beta \in [\beta_0,\beta_{max}]$. Now when $I=0$, $\lambda_1+\lambda_2=-d<0$ and when $I=b$
		\begin{align*}
			\lambda_1+\lambda_2(b)=\frac {1}{4}-d-\beta b
		\end{align*}
		Now we have
		\begin{align*}
			\beta &\le \beta_{max}\\
			\Rightarrow\frac {1}{4}-d-\beta b &\ge \frac {1}{4}-d-b\beta_{max} \\
			\Rightarrow\frac {1}{4}-d-\beta b& \ge\frac {1}{4} -d- \frac {(\delta+1)db}{\Delta}\\
			\Rightarrow\frac {1}{4}-d-\beta b &\ge \frac {1}{4} -d(1+ \frac {(\delta+1)b}{\Delta})>0\\
			\Rightarrow\lambda_1+\lambda_2(b)&=\frac {1}{4}-d-\beta b>0
		\end{align*}
	So under above assumptions, for each $\beta \in [\beta_0,\beta_{max}]$, we have $\lambda_1+\lambda_2(0)<0$ and $\lambda_1+\lambda_2(b)>0 $. So according to the \emph{intermediate value theorem}, for each $\beta \in [\beta_0,\beta_{max}]$ there is some $I_1\in (0,b)$ such that $\lambda_1+\lambda_2(I_1)=0$. But if we consider  $I_1$ as the intersection point of function $\frac {Ib}{(I+b)^2}$ and line $\beta I+d$, considering the behavior of this function and line, there is just one $I_1$ for each $\beta \in [\beta_0,\beta_{max}]$ and $I_1(\beta)$ is a continuous and $0<I_1(\beta)<b$ and $\lambda_1+\lambda_2(I_1(\beta))=0$ when $\beta \in [\beta_0,\beta_{max}]$.
	\\
	 Now we consider the quadratic equation \eqref{quad}. When $\frac{d}{(\delta+1)\beta}<b$ and $R_0=1$ or equally $\beta=\beta_0$, there is just one permissible solution, $\I=0$, and when $R_0>1$, there is always just one permissible root, which is a continuous function of $R_0$ or other parameters like $\beta\:(\I(\beta))$. Now we consider the root of quadratic equation \eqref{quad} for $\beta=\beta_{max}$. In this case, the quadratic equation \eqref{quad} will be expressed as follow:
	\begin{align*}
	P_M(I)=\delta I^2+(b(\delta+1)-A)I-bA+\frac {\delta d}{\beta_{max}}I+\frac {bd}{\beta_{max}}(\delta+1)
	\end{align*}
	Now we will consider the $P_M(b)$ and the condition $\frac{b(2\delta+1)}{2}<A$:
	\begin{align*}
		P_M(b)&=\delta b^2+(b(\delta+1)-A)b-bA+\frac {\delta db}{\beta_{max}}+\frac {bd}{\beta_{max}}(\delta+1)\\
		&=b(\delta b+b(\delta+1)-2 A)+b(\frac {\delta d}{\beta_{max}}+\frac {d}{\beta_{max}}(\delta+1))\\
		&=-2b(A-\frac{b(2 \delta+1)}{2})+\frac {b(2\delta+1)d}{\beta_{max}}\\
		&=b(-2 \Delta +\frac {(2\delta+1)\Delta}{\delta+1})\\
		&=b\Delta(-2+\frac {2\delta+1}{\delta+1})\\
		&=b\Delta(\frac {-1}{\delta+1})<0
	\end{align*}
	So obviously, $P_M(I)$ has a root $\I$ in the interval $(b, +\infty)$. In other words, with assumption of the conditions $\frac{d}{(\delta+1)\beta}<b$ and $\frac{b(2\delta+1)}{2}<A$, the quadratic equation \eqref{quad} has a root $\I=0$ when $\beta=\beta_0$ and a root $\I>b$ when $\beta=\beta_{max}$ and the $I^*(\beta)$ is a continuous function of $\beta \in [\beta_0,\beta_{max}]$. Now we use \emph{intermediate value theorem} for the following function $f(\beta)=I_1-I^*(\beta)$ and $\beta \in [\beta_0,\beta_{max}]$. This function is obviously continuous and 
	\begin{align*}
		&f(\beta_0)=I_1(\beta_0)-I^*(\beta_0)=I_1(\beta_0)>0\\
		&f(\beta_{max})=I_1(\beta_{max})-I^*(\beta_{max})<0\\
	\end{align*}
	So there is a $\beta_1 \in (\beta_0,\beta_{max})$ such that $f(\beta_1)=0$ or equally $I_1(\beta_1)=I^*(\beta_1)$ and 
	\begin{align*}
		\lambda_1+\lambda_2(I^*(\beta_1))=\lambda_1+\lambda_2(I_1(\beta_1))=0
	\end{align*}
	Now with considering the differentiability of functions $\I(\beta)$ and $\lambda_1+\lambda_2(I)$ and \emph{intermediate value theorem}, it will be possible to find a $\beta_1$ with the following properties:
	\begin{align*}
	(\exists \epsilon>0)\: \I(\beta_1)=I_1(\beta_1) \:,
		\begin{cases}
		\lambda_1+\lambda_2(\I(\beta))<0 \qquad \beta \in(\beta_1-\epsilon,\beta_1)\\\\
		\lambda_1+\lambda_2(\I(\beta))=0 \qquad \beta= \beta_1\\\\
		\lambda_1+\lambda_2(\I(\beta))>0 \qquad \beta \in(\beta_1,\beta_1+\epsilon)
		\end{cases}
	\end{align*}
	
	Now we will consider $\lambda_1\lambda_2(I)$. We have shown that $\lambda_1\lambda_2(I)$ will be positive when $I \in (0, +\infty)$ and $\frac{d}{(\delta+1)\beta}<b$. So:
	\begin{align*}
		\begin{cases}
		\lambda_1+\lambda_2(\I(\beta_1))=0\\\\
		\lambda_1\lambda_2(\I(\beta_1))>0\\
		\end{cases}	
		\Rightarrow -\lambda_1^2>0
		\Rightarrow
		\begin{cases}
		\lambda_1(\I(\beta_1))=i\omega_0\\\\
		\lambda_2(\I(\beta_1))=-i\omega_0
		\end{cases} 
		, \omega_0>0
	\end{align*}
	Now obviously, $\lambda_{1,2}$ are continuous function of $\I$ and $\beta$ and $\I$ is a continuous function of $\beta$, so $\lambda_{1,2}(\beta)$ will be continuous  with values in complex plane $\mathbb{C}$. So we can choose a $\epsilon$, which has been defined in the above arguments, with	 following properties:
	\begin{align*}
		Im(\lambda_{1}(\I(\beta)))>0
		 \qquad,\beta\in(\beta_1-\epsilon,\beta_1+\epsilon)
	\end{align*}
	
	Now because of $\lambda_2=\bar{\lambda}_1$, the above results can be summarized and rewritten as follow:
	\begin{align*}
		[(\exists \epsilon>0)\forall\beta\in(\beta_1-\epsilon&,\beta_1+\epsilon)] \:\lambda_{1,2}(\beta)=r\pm i\omega \qquad ,\omega>0\\\\
		,&\begin{cases}
		r<0 \qquad \beta\in(\beta_1-\epsilon,\beta_1)\\\\
		r=0  \qquad \beta=\beta_1\\\\
		r>0 \qquad \beta\in(\beta_1,\beta_1+\epsilon)
		\end{cases}
	\end{align*}
	
	So there must be a Hopf bifurcation when $\beta=\beta_1 \in (\beta_0,\beta_{max})$ or equally $R_0=\frac{\beta_1A}{(\delta+1)\beta} \in (1,\frac {A}{\Delta})$.$\Box$
	\end{Proof}
	
	As an example, this theorem can be seen in the following plot:
	\begin{center}
		\includegraphics[width=.7\linewidth]{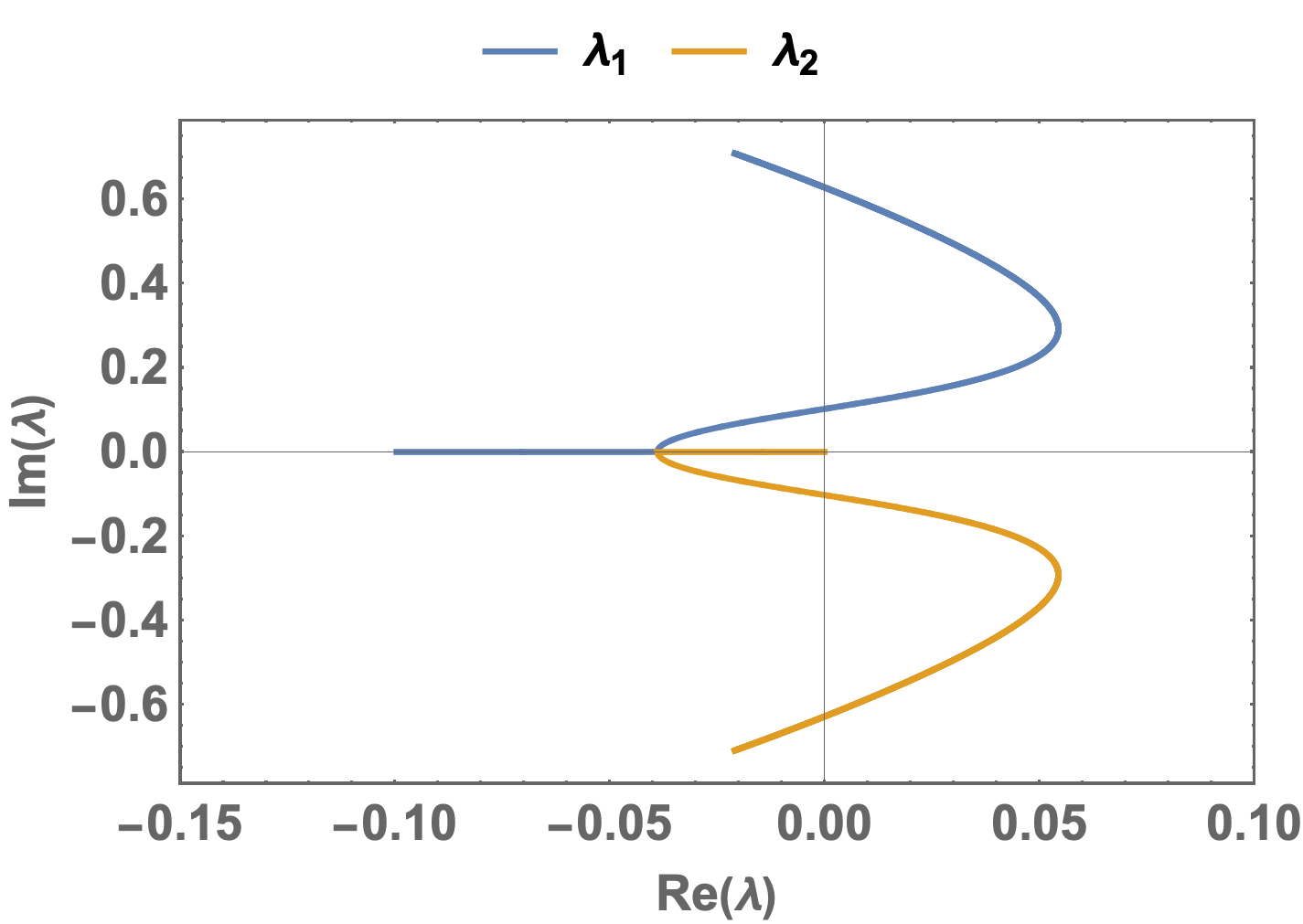}
	\end{center}
		\textbf{Figure \nfig:} Simulation of theorem 4.3: curves of eigenvalues in complex plane, where the blue and yellow curves indicate the different eigenvalues, and when the parameters are $A=10,d=.1,\alpha=1.9,n=1,b=1$ and $\beta$ changes in the interval $(.04,.09)$ or equally $R_0\in(1,2.25)$. The quantities and parameters have been defined in the tables 1 and 2 and $R_0$ has been defined in the section 2.\\
		
	It is easy to check that these parameters satisfy the conditions of this theorem and $\beta_0=.04$ and $\beta_{max}\thickapprox.06154$. Obviously, there are two Hopf bifurcations in this case. These bifurcations occur  $\beta_1\thickapprox0.040968$ and $\beta_2\thickapprox0.078668$ or equally $R_0^{(1)}\thickapprox1.0242$ and $R_0^{(2)}\thickapprox1.9667$. The case $\beta_1$ was predicted in the our theorem.
	\section{Numerical Analysis}
	\label{num}
	 In this section, we will concentrate on the regime of parameters where there is a Hopf bifurcation and analyze the time evolution of the dynamics. Thus for instance, we can consider the case in the Figure 5, where $\beta$ changes in the interval (.04,.09) and obviously, there is a limit cycle in this situation for $\beta\in(0.040968,0.078668)$. Now for a better observation of behavior of system, when $\beta$ changes, we will choose two different $\beta$, before (figures 6) and after (figures 7) $\beta_1\thickapprox0.040968$, with two different initial conditions for each $\beta$ and sketch stream plot $S-I$ and $S,I,H,R$ and fluxes as function of t, where the fluxes are defined as follow:  
	\begin{align} 
	\label{Eq.Flux}
	\begin{cases} 
	\phi_{in}:=\beta I S\\\\
	\phi_{out,1}:=(d+\alpha+n)I\\\\
	\phi_{out,2}:=\frac {bI}{I+b}
	\end{cases}
	\end{align}
	\\
	
	\begin{center}
		\begin{tabular}{ccc}
			\includegraphics[width=.3\linewidth]{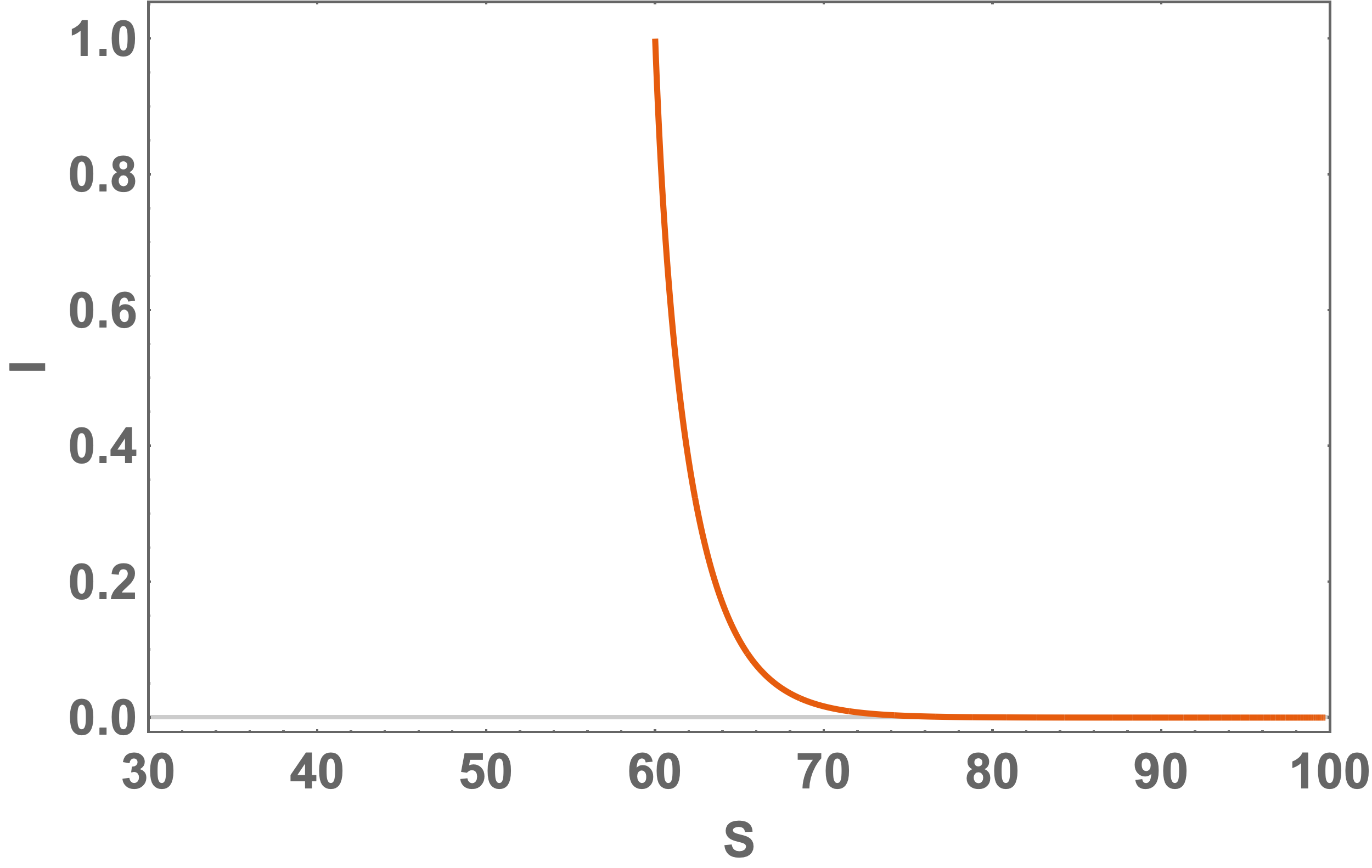}&
			\includegraphics[width=.3\linewidth]{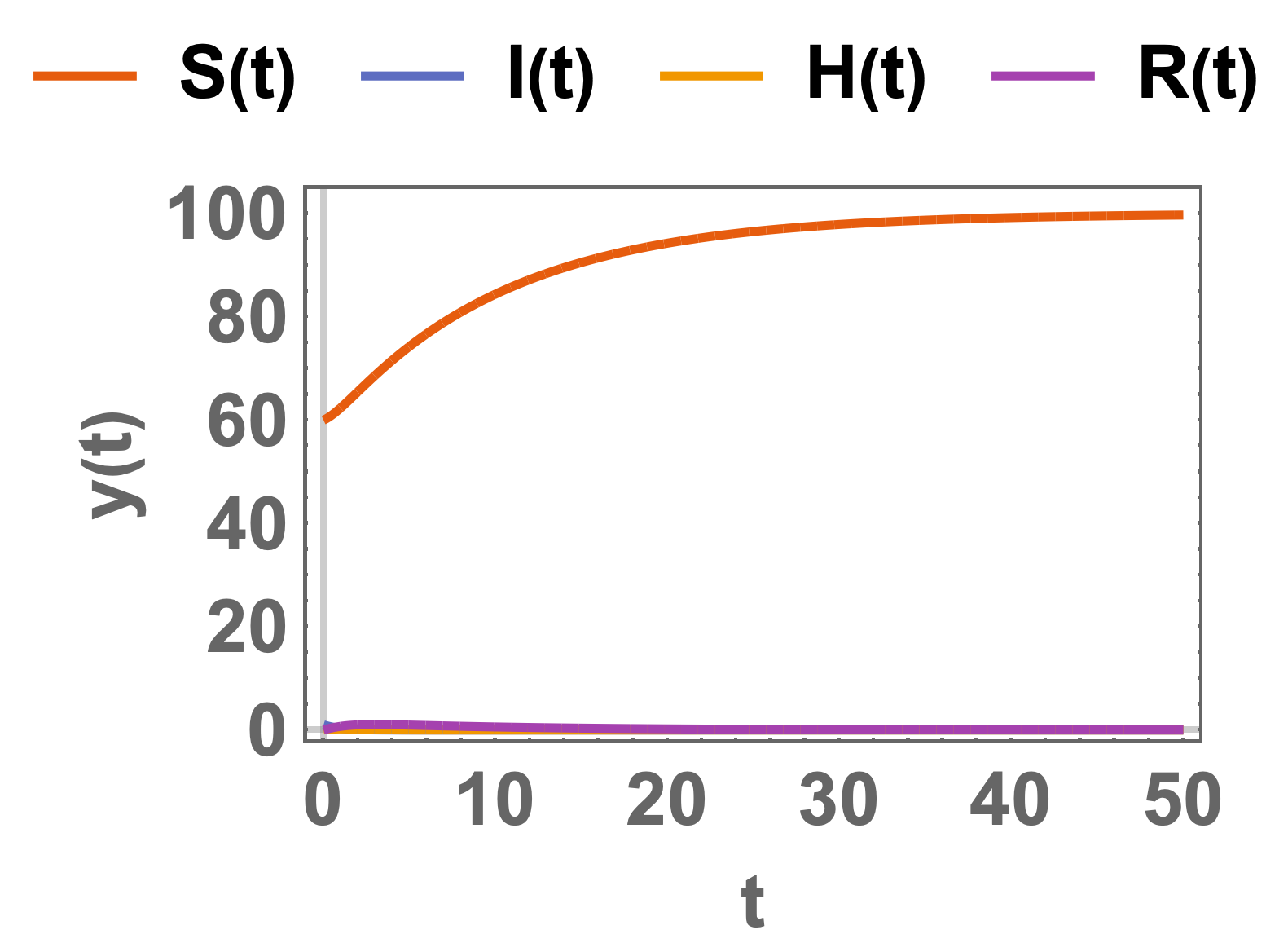}&
			\includegraphics[width=.3\linewidth]{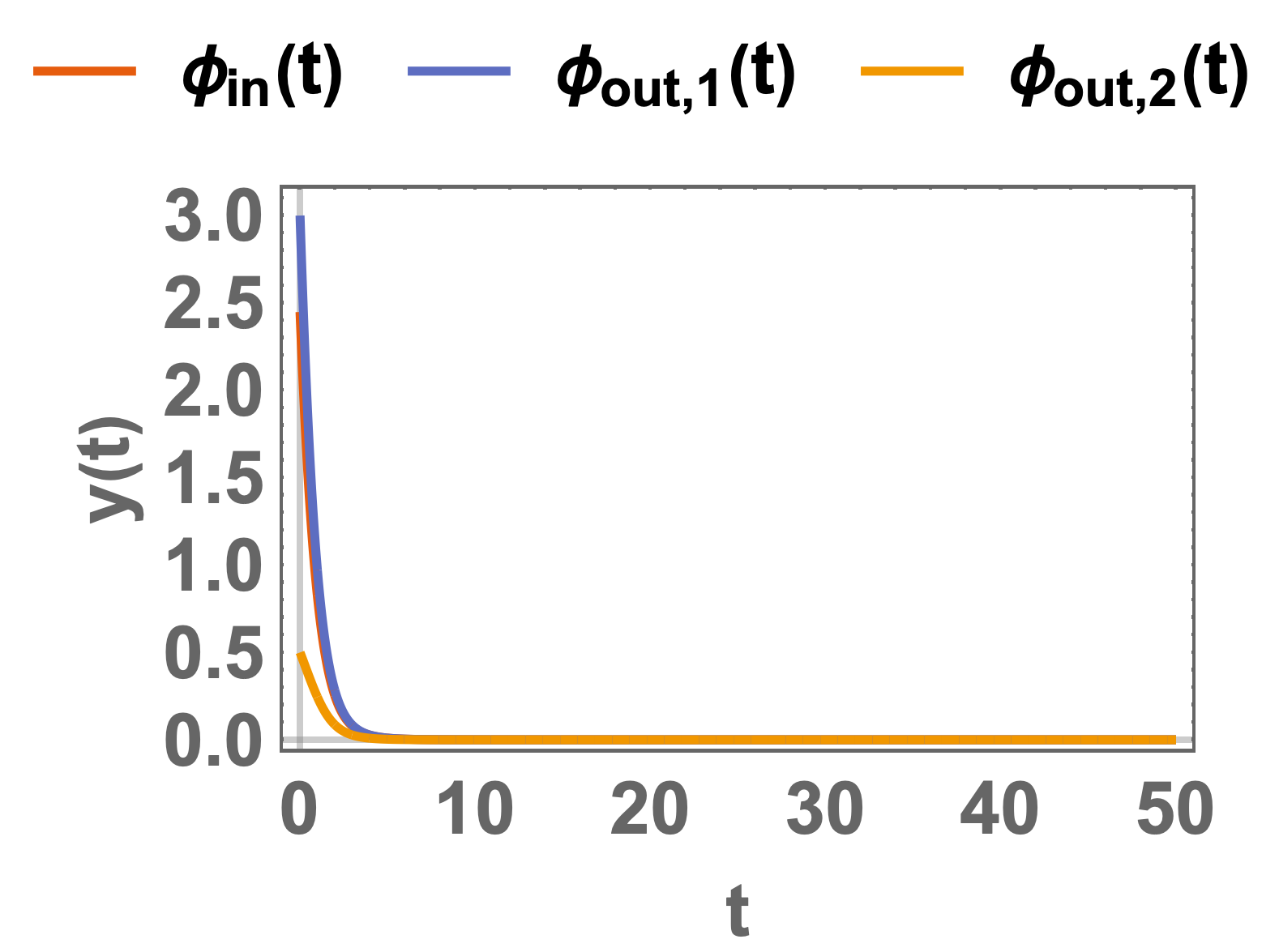}\\
			$(A)$&$(B)$&$(C)$\\
    		\includegraphics[width=.3\linewidth]{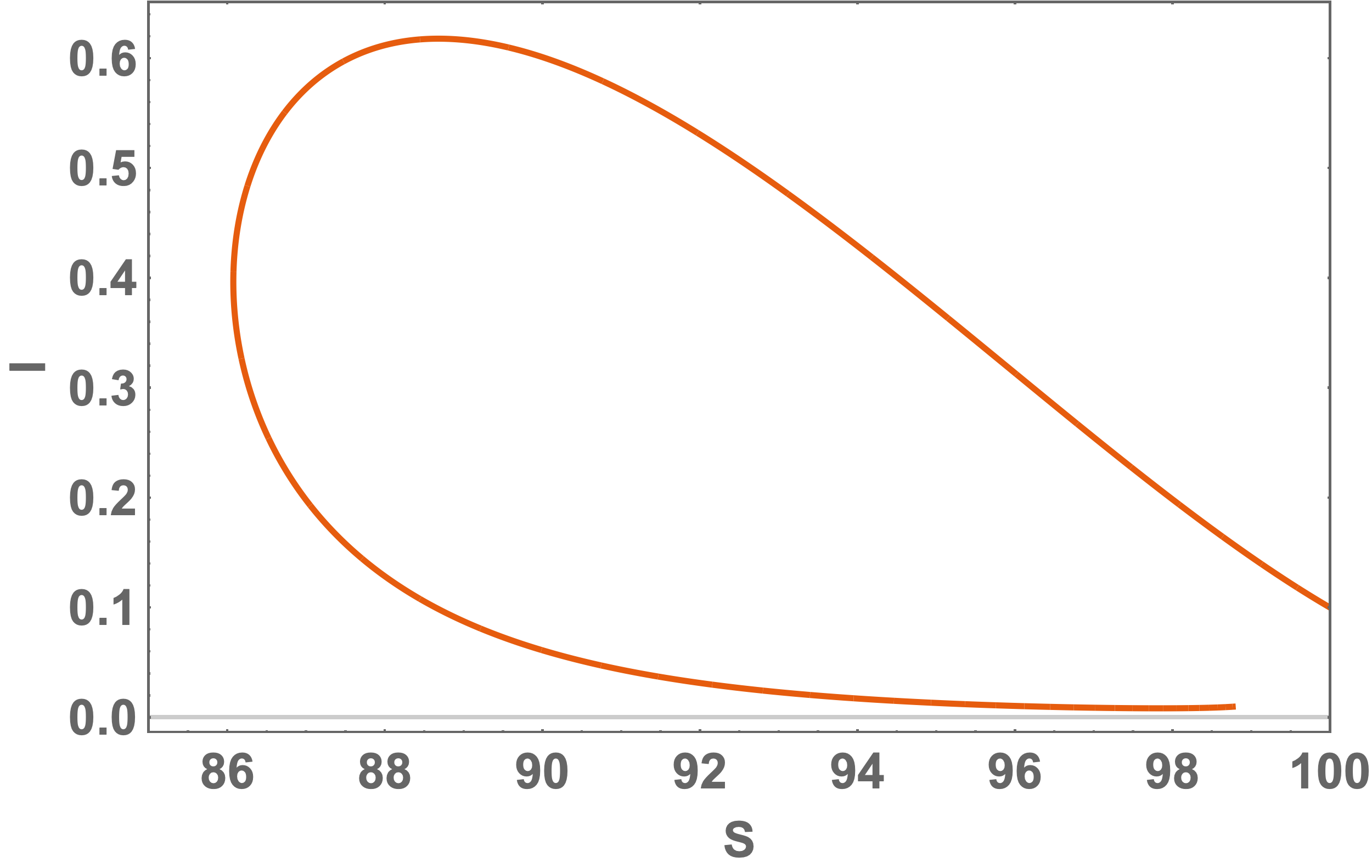}&
    		\includegraphics[width=.3\linewidth]{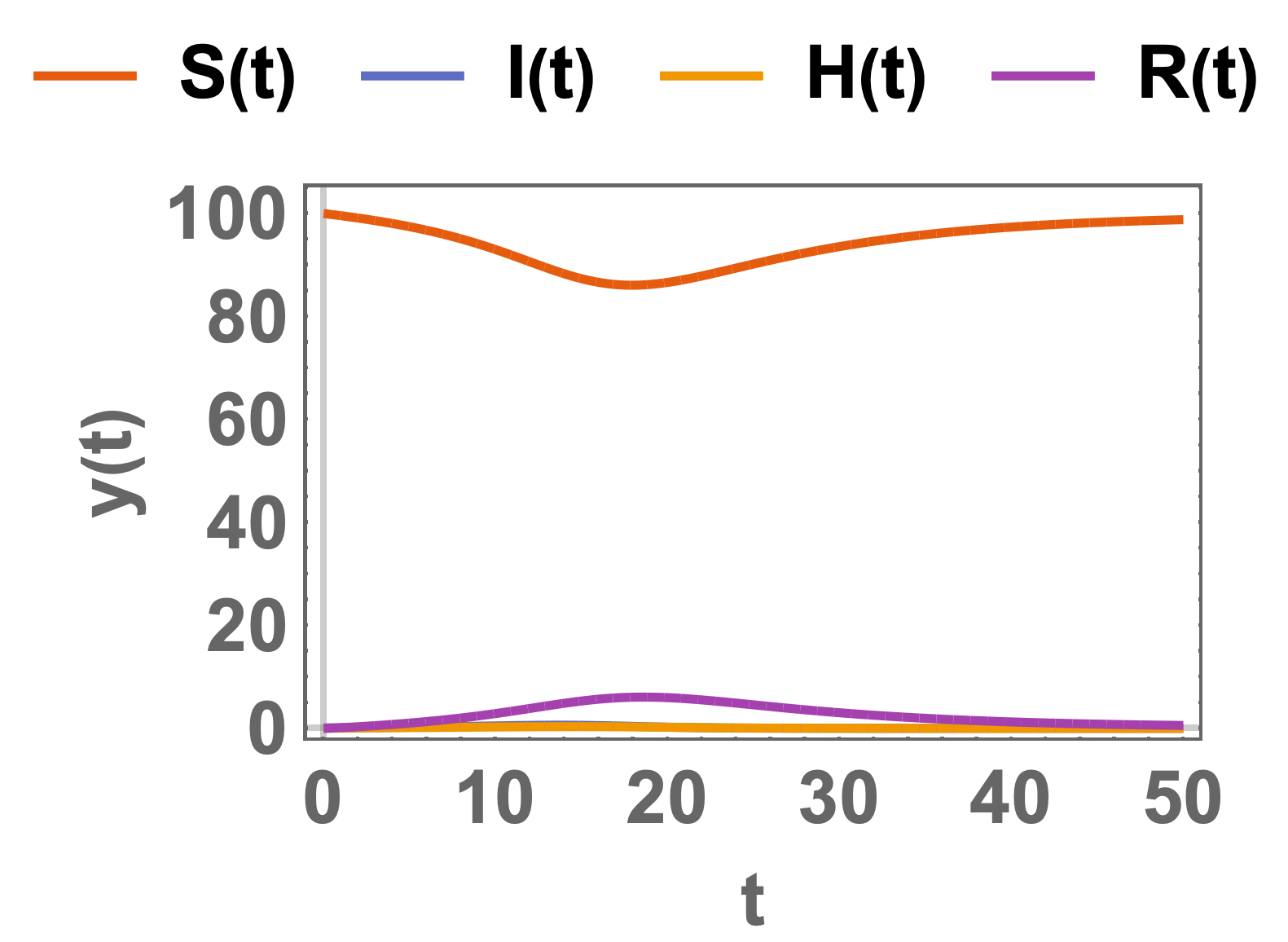}&
    		\includegraphics[width=.3\linewidth]{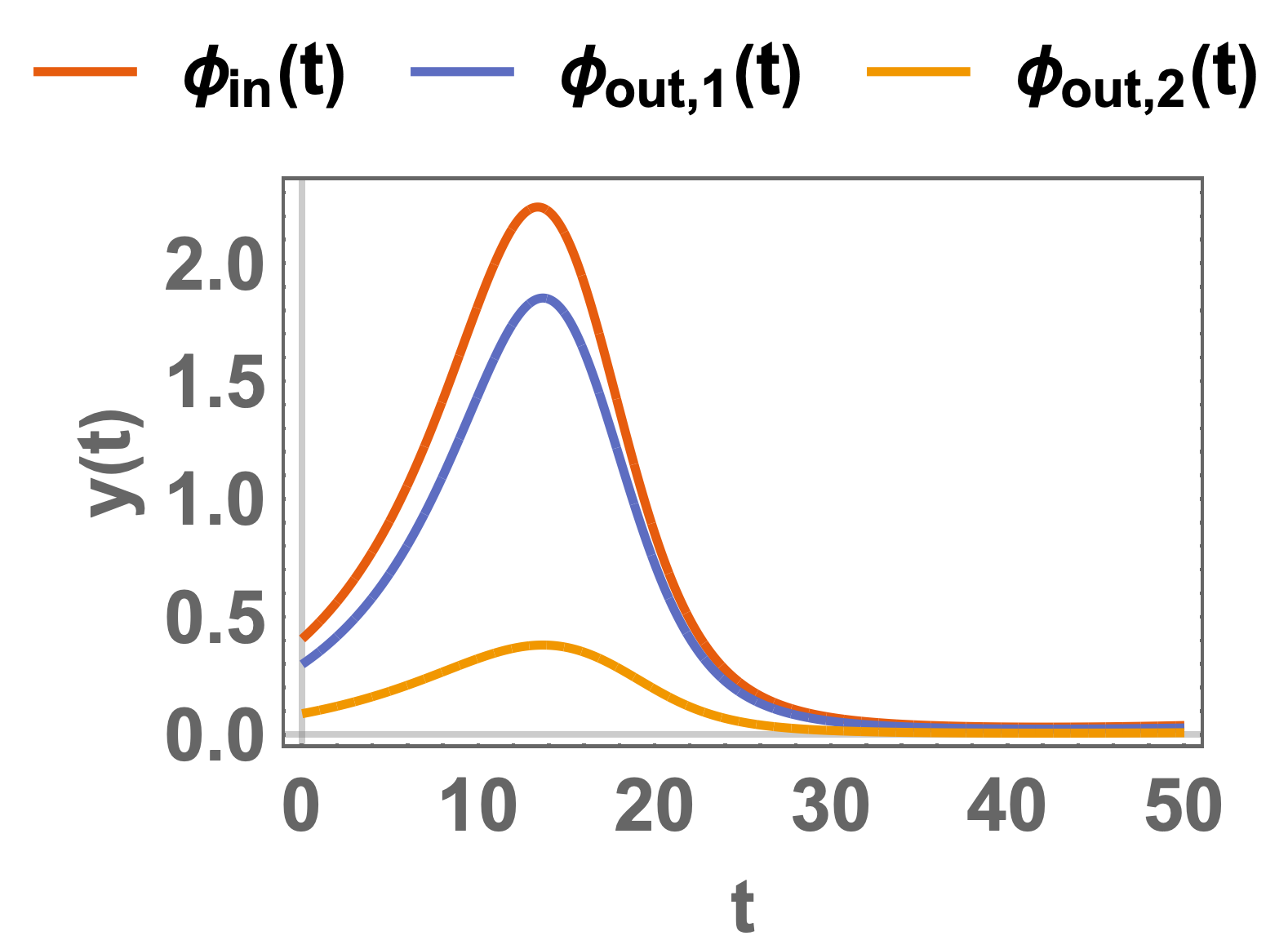}\\
    		$(A')$&$(B')$&$(C')$\\
		\end{tabular}
	\end{center}
		\textbf{Figure \nfig:} stream plot $S-I$ $(A,A')$ and plots of $S,I,H,R$ $(B,B')$ and fluxes $(C,C')$ as  a function of $t$ when the parameters are $b=1,A=10,d=.1,\alpha=1.9,n=1,n'=1$ and $\beta=.0408$ and initial conditions are $X_0=(60,1,0,0)$ For panels $(A,B,C)$ and $X_0=(100,.1,0,0)$ $(A',B',C')$. The quantities and parameters have been defined in the table 1 and 2 and the fluxes have been defined in the equations \ref{Eq.Flux}.
	
	\begin{center}
		\begin{tabular}{ccc}
		\includegraphics[width=.3\linewidth]{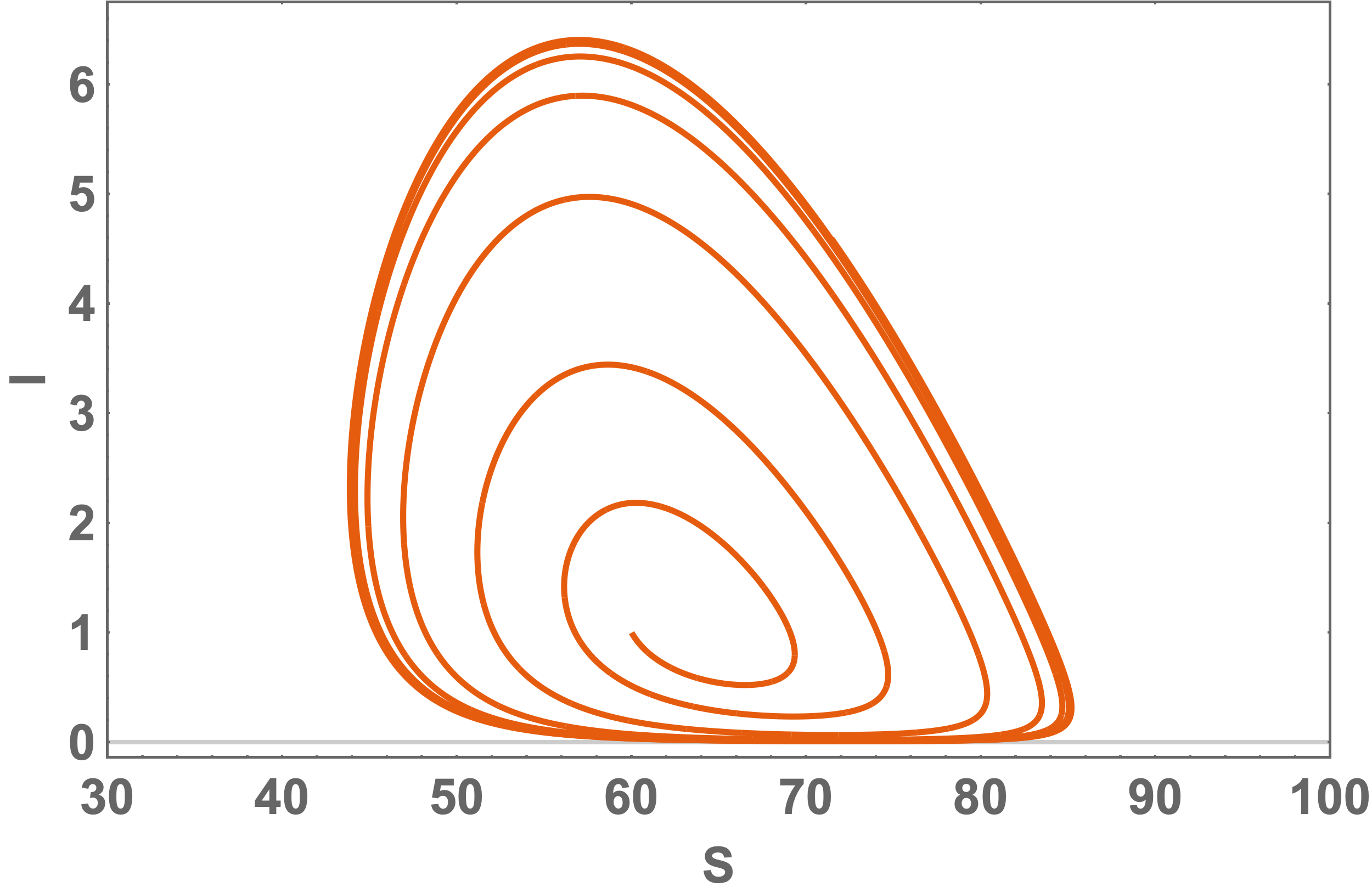}&
		\includegraphics[width=.3\linewidth]{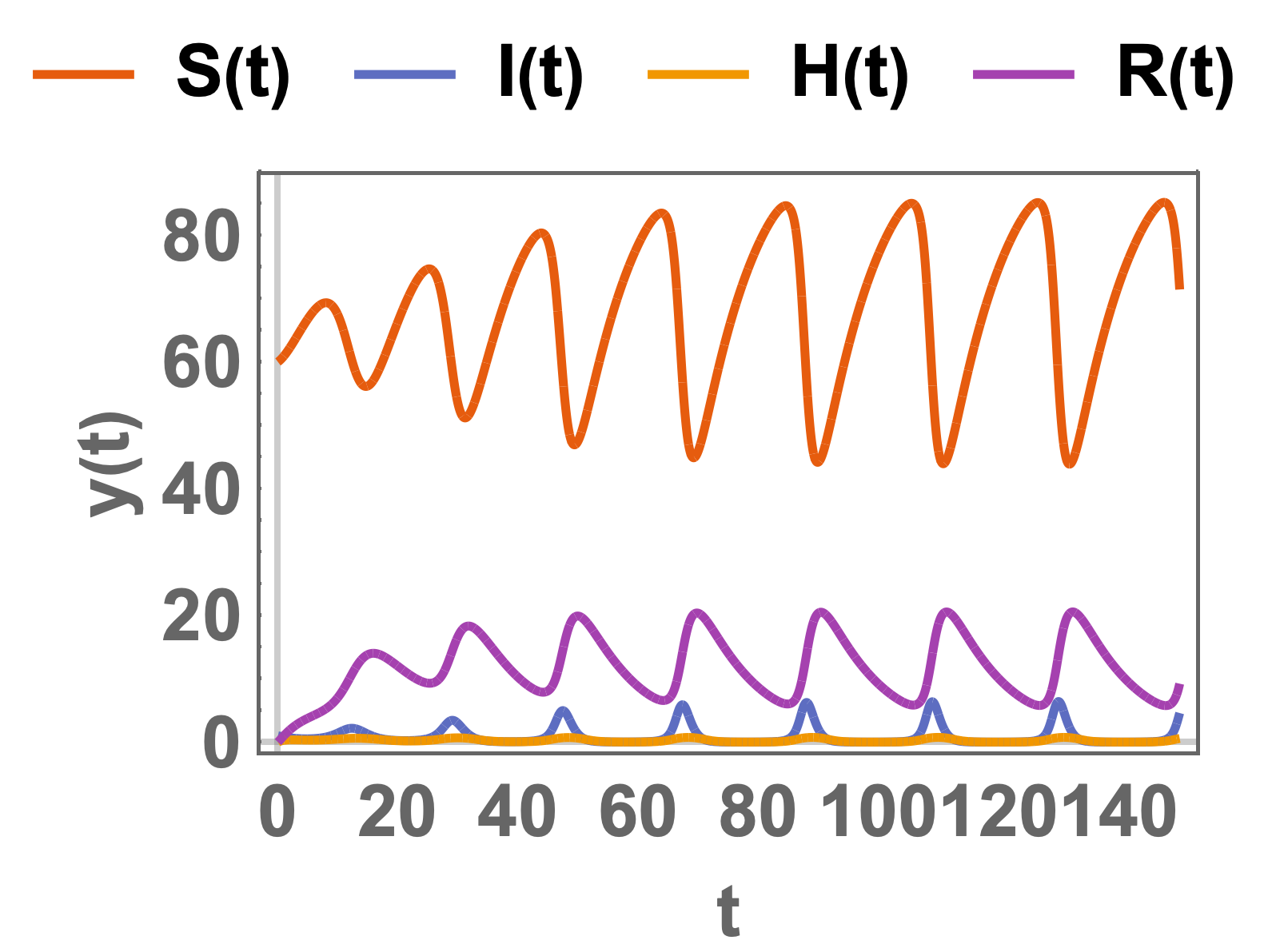}&
		\includegraphics[width=.3\linewidth]{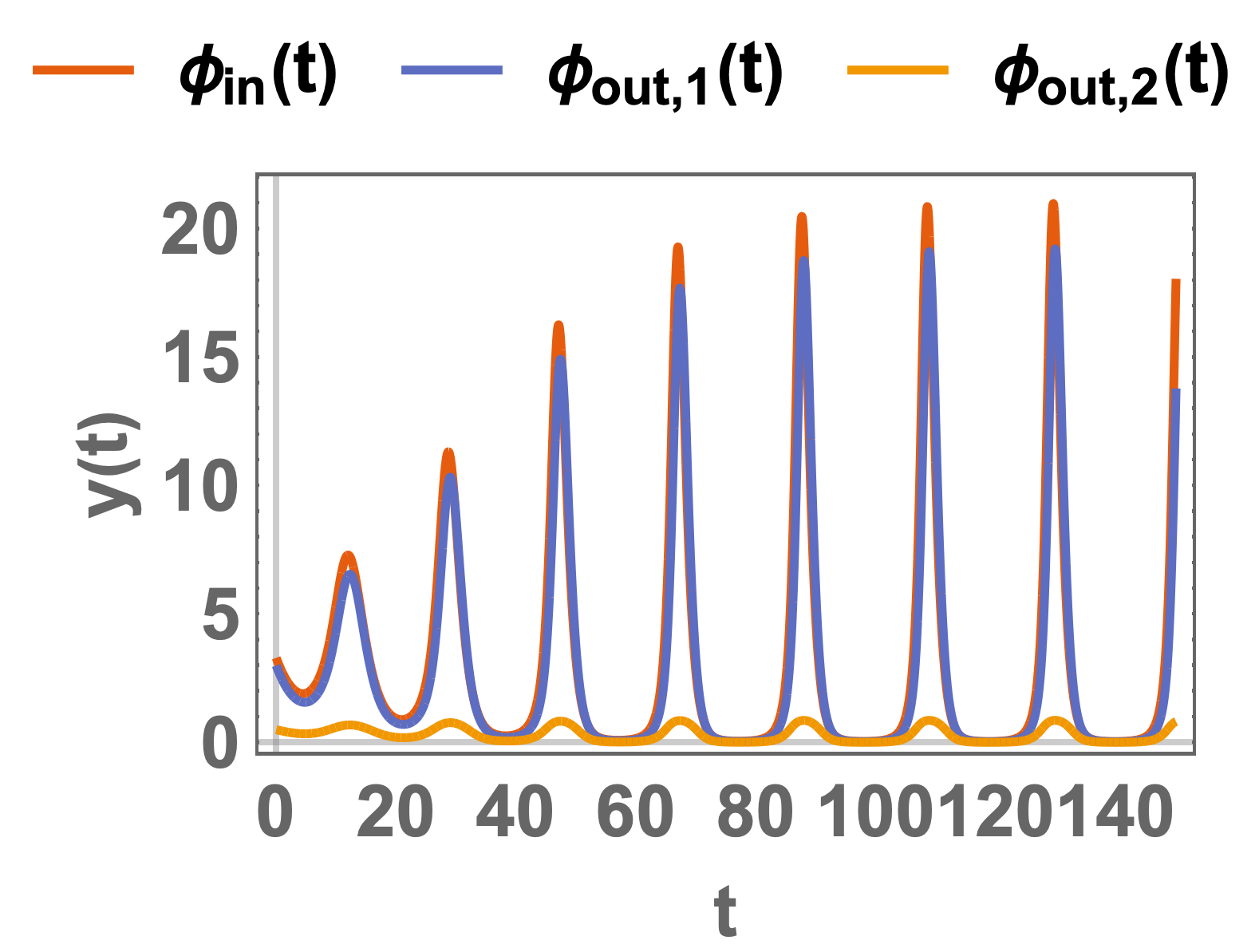}\\
		$(A)$&$(B)$&$(C)$\\
		\includegraphics[width=.3\linewidth]{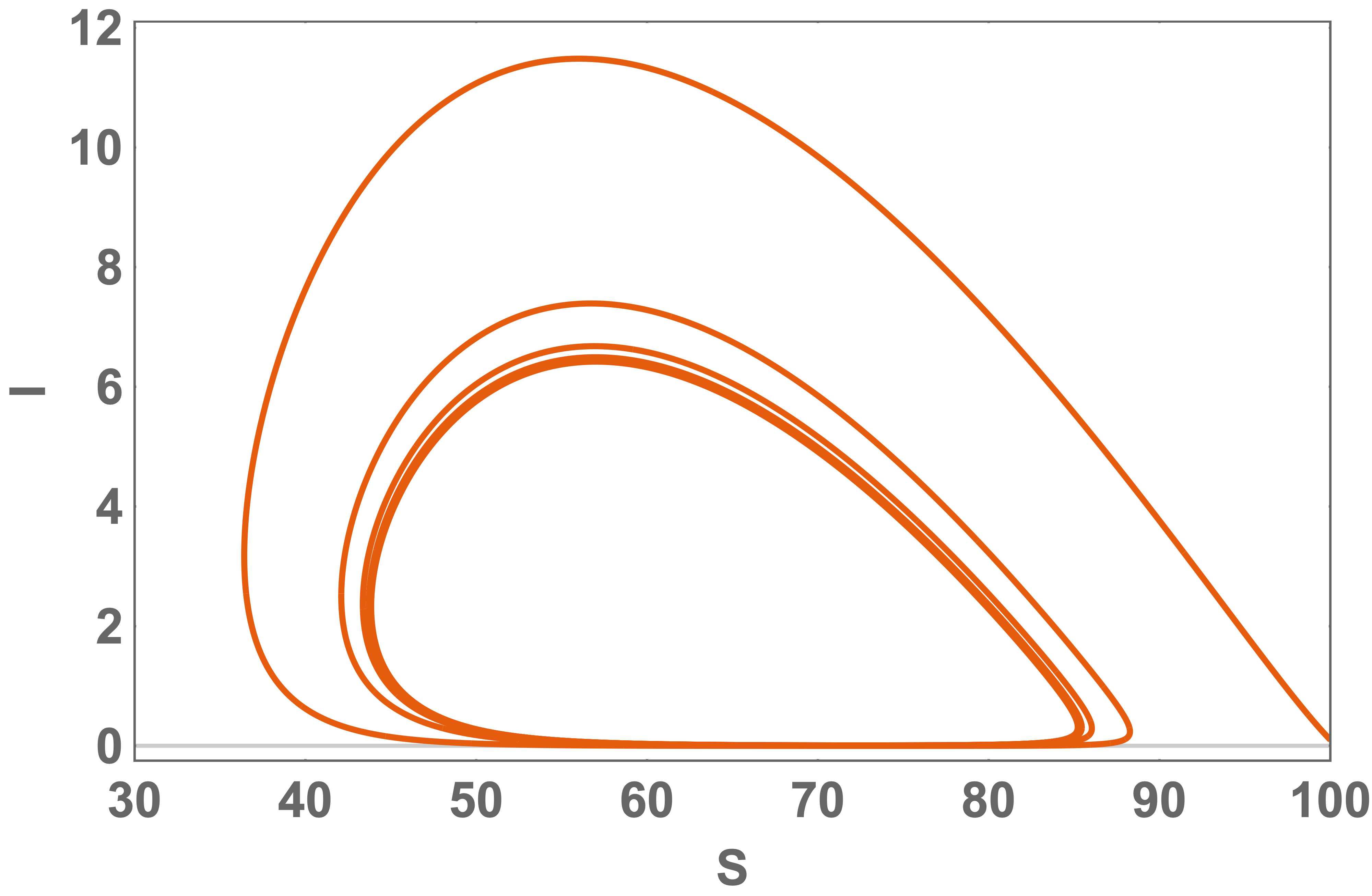}&
		\includegraphics[width=.3\linewidth]{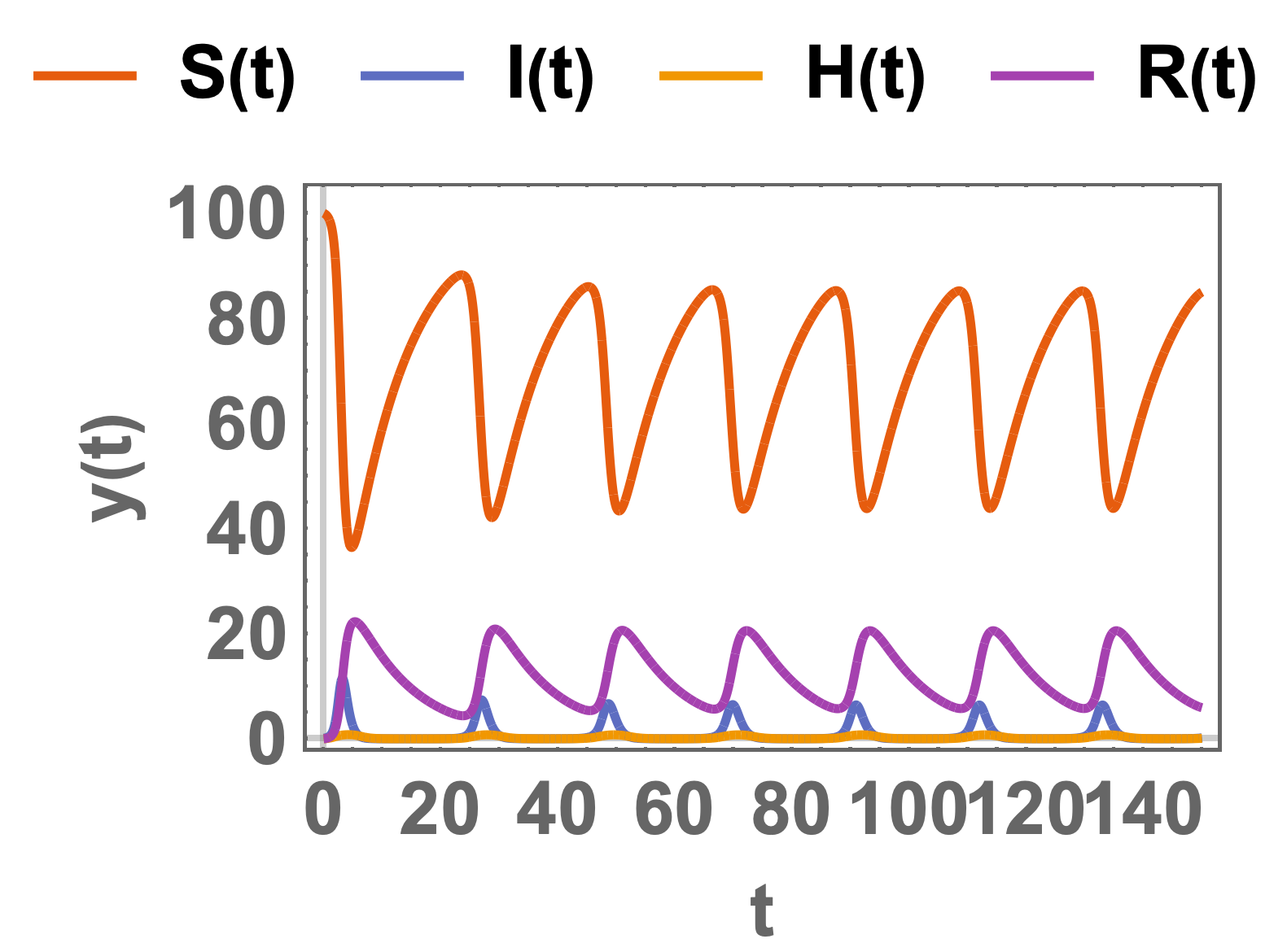}&
		\includegraphics[width=.3\linewidth]{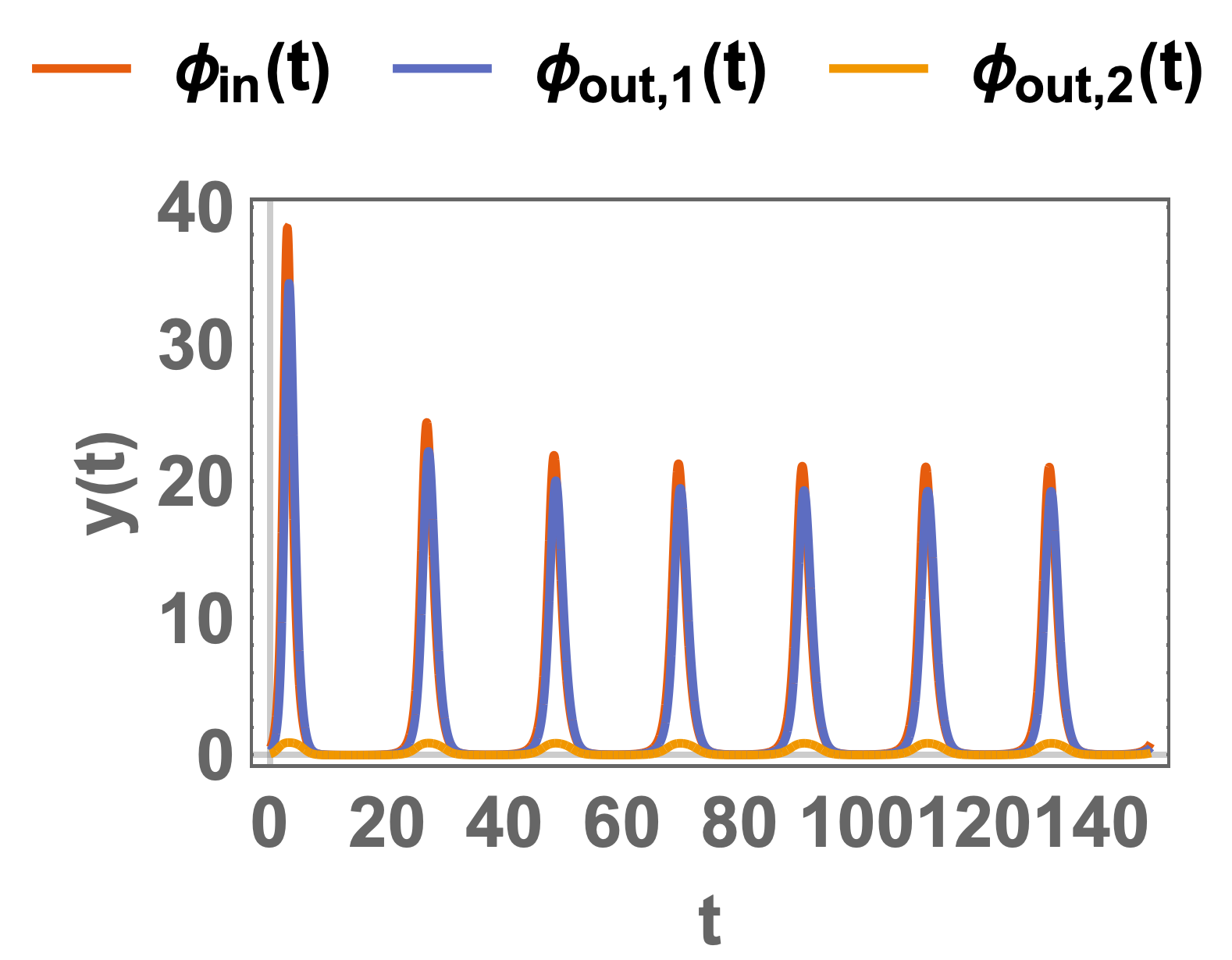}\\
		$(A')$&$(B')$&$(C')$\\
		\end{tabular}
	\end{center}
		\textbf{Figure \nfig:} stream plot $S-I$ $(A)$ and plots of $S,I,H,R$ $(B)$ and fluxes $(C)$ as a function of $t$ when the parameters are $b=1,A=10,d=.1,\alpha=1.9,n=1,n'=1$ and $\beta=.055$ and initial conditions are $X_0=(60,1,0,0)$ for panels $(A,B,C)$ and $X_0=(100,.1,0,0)$ for panels $(A',B',C')$.The quantities and parameters have been defined in the table 1 and 2 and the fluxes have been defined in the equations \ref{Eq.Flux}.\\

	In the figure 6, $\beta$ is less than 0.040968, so there is no limit cycle in this condition and curves in the stream plots (panels $(A,A')$) approach to a fixed point. Because of that, plots of $S,I,H,R$ and fluxes have limit when $t\rightarrow\infty$. In contrast, in the figure 7, $\beta>0.040968$ and there is a limit cycle and curves in the stream plots (panels $(A,A')$) approach to this limit cycle. Because of that, plots of $S,I,H,R$ and fluxes behave periodic when $t$ is large enough. 
	
	\section{Discussion}
	In our study, we changed the standard SIR model and divided survived infected individuals into block $R$ and a new introduced block $H$ in order to model number of individuals of ICU, who need special and intensive treatment, where hospitalization rates were defined as $nI$ and $\frac{bI}{I+b}$ respectively. Our model can be considered as a special case of model by Zang et al. \cite{zhang2008backward} but in our study we have concentrated on existence and type of bifurcations, which is a new aspect of previous model, and we could observe forward, backward and Hopf bifurcation.  
	
	The most important result of our study, which makes our study different from previous works, is existence of Hopf bifurcation and so existence of limit cycle. Existence of limit cycle shows that a disease can survive in the society and this is a difficult and critic situation for governments in order to control this disease.
	
	As an application of our model, we will consider following case, which has been based on existence of Hopf bifurcation. Suppose that we are in a regime of parameters where a supercritical Hopf bifurcation can occur when $\beta=\beta_0$ like the case, which has been considered in numerical analysis section \ref{num}. After spread of a special disease, governments usually use interventions in order to reduce contact between people for example quarantine. So $\beta$ is small and we can suppose that $\beta<\beta_0$, where there is no limit cycle. But after a while, it is possible that $\beta$ increases and $\beta>\beta_0$. This event can happen for example when people ignore the interventions of governments and quarantine. And in this situation, there is a limit cycle and disease can survive.
	
	As a limitation of our study, we can mention that we could not analyze time evolution of fluxes, which have defined in the equations \eqref{Eq.Flux}, although we have simulated them numerically in the section \ref{num}. Please note that the fluxes are important since they are the practical observable in the hospitals and can be used as a sort of warning signal, also can help us to estimate the other parameters of the dynamics. Therefore further theoretical works are needed in order to figure out how to detect such behaviours in practice. 
	
	\newpage
	\printbibliography[heading=bibintoc]
\end{document}